\documentclass[aps,prl,preprintnumbers,superscriptaddress,longbibliography,amsmath,amssymb,twocolumn,10pt,nofootinbib]{revtex4-2}
\usepackage{newtxtext,newtxmath}
\usepackage{subfigure}
\usepackage{wrapfig}
\usepackage{enumitem}
\usepackage{xspace}
\usepackage{amsmath}
\usepackage{amsfonts} 
\usepackage{appendix}
\usepackage{graphicx}% Include figure files
\usepackage{dcolumn}% Align table columns on decimal point
\usepackage{bm}% bold math
\usepackage{color}
\usepackage{xcolor}
\usepackage[colorlinks=true]{hyperref}

\begin{document}

\preprint{AUGER-2026-03}

\title{Bounds on Lorentz invariance violation from muon fluctuations at the Pierre Auger Observatory}
\author{The Pierre Auger Collaboration}
\affiliation{The Pierre Auger Observatory, Av.\ San Mart\'in Norte 306, 5613 Malarg\"ue, Mendoza, Argentina;
\\
http://www.auger.org}
\email{spokespersons@auger.org}

\date{\today}

\begin{abstract}
  Quantum gravity theories often modify spacetime symmetries. In particular, Lorentz invariance may be violated when approaching the Planck scale. Although the scales at which interactions occur in extensive air showers induced by ultra-high-energy cosmic rays in the atmosphere are many orders of magnitude below the Planck scale, these violations might still be observable. In this work, the fluctuations in the number of muons in the extensive air showers measured at the Pierre Auger Observatory are exploited, for the first time, to constrain Lorentz invariance violations. The bounds derived in the hadronic sector are the strongest ever obtained, and do not rely on assumptions about the mass composition of ultra-high-energy cosmic rays.
  The fluctuations in the number of muons constitute a new and powerful observable to further explore Lorentz invariance in a region of the parameter space not accessible to other observables. 

\end{abstract}

\maketitle

\bibliographystyle{apsrev4-1} 

The effort for a unified theory of gravity has led to the development of several quantum gravity (QG) models, some of them implying the breaking of Lorentz invariance (LI)~\cite{Aloisio:2000cm}. While this would appear at the scale of the Planck mass, the highest energetic processes in our Universe, namely those involving the interactions of ultra-high-energy cosmic rays (UHECRs), could keep track of this breaking. During their propagation in extragalactic space, UHECRs undergo interactions with background photons such as meson photoproduction or photodisintegration; Lorentz invariance violations (LIV) could affect their energy thresholds, suppressing such interactions and making the attenuation lengths extremely large~\cite{Aloisio:2000cm,Stecker:2009hj,Maccione:2009ju,Saveliev:2011vw}.
As a consequence, the existing evidence of the suppression of the UHECR flux at the highest energies~\cite{PierreAuger:2020qqz,PierreAuger:2025eun} can be exploited to set limits on LIV. 
In particular, LIV can be tested by searching for the best description of UHECR observables in terms of astrophysical scenarios, under LIV assumptions, as already attempted for instance in~\cite{Bi:2008yx,Scully:2008jp,Boncioli:2015cqa,Boncioli:2017nec,Lang:2020geh,PierreAuger:2021tog}. However, the scenario is complicated by the fact that the best astrophysical description of the UHECR energy spectrum and mass composition is found corresponding to values of maximum energy of UHECRs at the sources smaller than or comparable to the typical threshold energy for photo-meson or photo-disintegration reactions~\cite{PierreAuger:2016use,PierreAuger:2022atd}. For this reason, the sensitivity to deviations from LI in the UHECR propagation is milder than expected, and alternative approaches need to be investigated. Independently of propagation effects, Lorentz invariance violation may also manifest itself in particle interactions at extreme energies, potentially altering the development of extensive air showers in the atmosphere.
Depending on the strength of the LI violation, the high energies available in the collision of cosmic rays with the atmosphere~\cite{footnote1} can lead to modifications of the shower development with respect to the standard LI case~\cite{Boncioli:2015cqa}, which is what is investigated in this work. The Pierre Auger Observatory~\cite{PierreAuger:2015eyc}, located near Malarg\"ue, in the Province of Mendoza in Argentina, is the largest cosmic-ray detector ever constructed, covering an area of 3000$~\mathrm{km}^2$. Completed in 2008, it serves as the most precise facility for UHECR detection above 0.1 EeV.
The basic design of the observatory comprises two components: the surface detector (SD) array, featuring 1660 water Cherenkov detectors (WCD) arranged in a triangular grid, and the fluorescence detector (FD), a set of telescopes overlooking the SD area. The SD, designed for nearly 100\% duty cycle operation, detects secondary particles at the ground level, while the FD, operational on clear moonless nights with a $\sim$15\% duty cycle, captures UV fluorescence light emitted by nitrogen molecules along the path of extensive air showers (EAS). 
This hybrid configuration guarantees a thorough comprehension of UHECRs by combining a detailed calorimetric measurement of the longitudinal profile with the FD, and assessing the lateral distribution of particles at ground through the SD. In 2015, the Pierre Auger Collaboration initiated a crucial upgrade phase known as AugerPrime~\cite{Schmidt:2025fnr}. The upgrade significantly improves the observatory's ability to discriminate muonic and electromagnetic components of showers on an event-by-event basis, offering valuable insights into UHECR physics. 
The Pierre Auger Observatory can measure the fluctuations in the number of muons using a subset of inclined hybrid events ($62^{\circ}< \theta < 80^{\circ}$) reconstructed simultaneously by the SD and FD, for which the pure electromagnetic component is largely absorbed in the atmosphere and the particles reaching the ground directly sample the muon content. The measurement used in this work is based on data
collected between January 1, 2004 and December 31, 2017; after the full event
selection, 786 events remain, of which 281 with energy $\textrm{E} > 4 \times 10^{18} \textrm{eV}$ are used to extract
the fluctuations, thus ensuring the full trigger efficiency for the SD. Within the current uncertainties, an agreement between the experimental relative fluctuations and the models has been found~\cite{PierreAuger:2021qsd}.
In this context, the event-by-event fluctuations in the number of muons at ground provide a particularly sensitive probe, as they are dominated by the first hadronic interactions and are less affected by the overall normalization and mass composition uncertainties.
In this work, for the first time, measurements of the relative fluctuations in the number of muons~\cite{PierreAuger:2021mve} in extensive air showers are exploited to constrain LIV in the hadronic sector, in a region of parameter space not accessible to the LIV modifications in the extragalactic propagation of cosmic rays.

\section*{\label{par:LIV}Lorentz invariance violation framework}

A well established phenomenological approach to introduce LIV effects consists of adding effective terms in the dispersion relation of a particle, arising from a metric modification (see~\cite{Aloisio:2000cm} for details). The dispersion relation would therefore appear as:
\begin{equation}\label{eq.1}
E^2 -p^2=m^2+f(\vec{p},M_{\text{Pl}}),
\end{equation}
where $m$ is the particle mass, $E$ its energy, $\vec{p}$ its momentum and $M_{\text{Pl}}$ is the Planck mass~\cite{footnote2}. At $\left|\vec{p}\right| \ll M_{\text{Pl}}$, $f$ can be expanded in series 
and considering only the leading order of the expansion, $n$, Eq.~\ref{eq.1} becomes:
\begin{equation}\label{eq.2}
E^{2}-p^{2}=m^{2}+ \eta^{(n)}\frac{p^{n+2}}{M_{\text{Pl}}^{n}},
\end{equation}
where $\eta$ is the dimensionless constant coefficient to be constrained.
 
The LIV hadronic sector is investigated through the expected cosmic-ray flux at Earth, compared to the measured energy spectrum and mass composition. Negative values of $\eta$ do not imply significant changes in the propagation of UHECRs~\cite{Boncioli:2015cqa,Boncioli:2017nec}; in~\cite{PierreAuger:2021tog}, positive values of $\eta$ are explored setting upper limits for different orders of violation, achieved through the flux suppression observed at the highest energies:  $\eta_{\pi}^{(1)}<2.4\cdot10^{-10}$ and $\eta_{\pi}^{(2)}<0.3$, at 5$\sigma$ C.L~\cite{footnote3}. The second order LIV parameter in the hadronic sector is also investigated in previous works, such as~\cite{Maccione:2009ju} with similar conclusions as in~\cite{PierreAuger:2021tog} although with assumptions on the UHECR mass composition~\cite{footnote4}.
Interpreting the right-hand side of the Eq.~\ref{eq.2} as an energy dependent mass: 
\begin{equation}\label{eq.LIVmass}m_{\text{LIV}}^2=m^{2}+\eta^{(n)}\frac{p^{n+2}}{M_{\text{Pl}}^n},
\end{equation}
the Lorentz factor for a LI violating particle at energy $E$ can be defined as:
\begin{equation}\label{eq.gamma}
\gamma^{~}_{\text{LIV}}=E/m^{~}_{\text{LIV}}.
\end{equation}

\begin{figure}[t]
 \centering
   \includegraphics[width=1.\columnwidth]{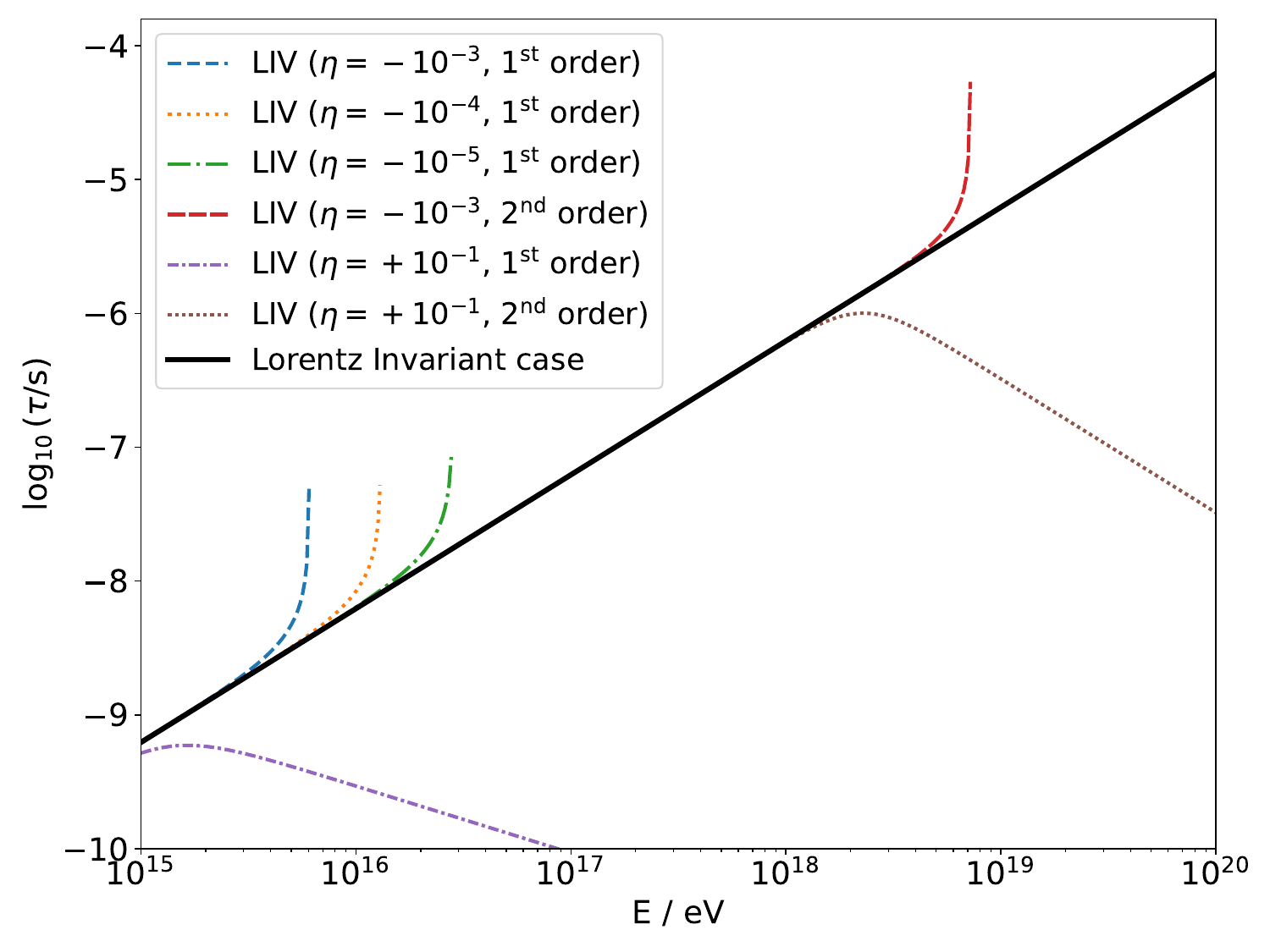}
 \caption{\label{fig:tau} Mean lifetime of the neutral pion as a function of the energy for the Lorentz invariant case and for different strengths and orders of LIV.}
\end{figure}

Depending on the value assumed by $\eta^{(n)}$, the lifetime of the considered particle $\tau=\gamma^{~}_{\text{LIV}}\tau^{~}_0$ will change accordingly. For negative/positive values of $\eta^{(n)}$, the lifetime of the particle should increase/decrease with respect to the LI case producing modifications in the EAS development. As an example, the $\pi^0$ lifetime as a function of the energy for the standard case and for different values of the LIV parameter is shown in Fig.~\ref{fig:tau}.

For negative values of $\eta$, the mean lifetime increases up to a critical energy, corresponding to the point at which the phase space reduces to zero (i.e. $m^2_{\text{LIV}} \rightarrow 0$) and the particle becomes stable (i.e. $\gamma \rightarrow \infty$). The energy at which the lifetime evolution deviates from the LI case depends both on the order and the strength of the violation. 
On the other hand, for positive values of the eta parameter the lifetime becomes smaller with respect to the standard LI case as the energy increases. Since neutral pions already decay promptly in the Lorentz-invariant case, the corresponding effects are expected to be much smaller for the observable considered here. For this reason, in this work we only focus on negative values of the $\eta$ parameter.
In order to describe the modifications in the development of a shower, let us first consider the simple model~\cite{Matthews:2005sd}, where a primary hadron interacts in the atmosphere, transferring a fraction $\alpha_{\rm {had},1}$ of its energy to hadronic particles (mainly charged pions, $\pi^{\pm}$), while the remaining $(1-\alpha_{\rm {had}, 1})$ fraction goes to neutral pions, $\pi^0$. Charged pions further interact, developing a hadronic cascade, while neutral pions decay promptly into photons, forming an electromagnetic sub-shower. The number of charged pions grows with each generation $i$ until energy depletion. At the critical generation $i=c$, charged pions predominantly decay into muons, yielding:
\begin{equation}
N_\mu = \frac{E_0}{\xi_c} \prod_{i=1}^{c} \alpha_{\mathrm{had},i},
\end{equation}
where $E_0$ is the primary energy and $\xi_c$ the pion critical energy. Fluctuations in $N_\mu$ arise from variations in $\alpha_{\mathrm {had},i}$:
\begin{equation}
\left(\frac{\sigma_{N_\mu}}{\langle N_\mu \rangle}\right)^2 = \sum_{i=1}^c \left(\frac{\sigma(\alpha_{\mathrm{had},i})}{\langle \alpha_{\mathrm {had},i} \rangle}\right)^2.
\label{eq.sigmaNmu}
\end{equation}

For large $i$, $\sigma(\alpha_{\mathrm{had},i})$ decreases as $\alpha_{\mathrm{had},i}$ averages over many interactions, making $\sigma(\alpha_{\mathrm{had},1})$ dominant~\cite{Cazon:2018gww}. In the LIV scenario with negative $\eta^{(n)}$, the $\pi^0$ lifetime increases, enhancing interaction probability before decay. This causes neutral pions to initiate hadronic sub-showers. As $\alpha_{\mathrm{had},j}^{\text{LIV}} \simeq 1$ for affected generations, fluctuations diminish ($\sigma(\alpha_{\mathrm{had},j}^{\text{LIV}}) \simeq 0$) and only LI generations contribute to fluctuations in Eq.~\eqref{eq:RelativeFluct}.  The number of muons, on the other hand, increases by $\prod_{j=1} (\alpha^{\text{LIV}}_{\mathrm{had},j} - \alpha_{\mathrm{had},j})$. In summary, LIV leads to a moderate muon excess and reduced fluctuations due to limited stochastic leakage in the first interaction~\cite{Cazon:2018gww}.

Since less energy is transferred to the electromagnetic component, the shower maximum depth ($X_{\text{max}}$~\cite{footnote5}) is slightly modified, and a displacement in the position of the maximum of the longitudinal profile is expected, together with a reduction of the maximum energy deposit in the atmosphere (see Appendix). 

\section*{\label{par:Muon}Simulating LIV effects in the shower development}
%simulation
\begin{figure}[t]
 \centering
   \includegraphics[width=1.\columnwidth]{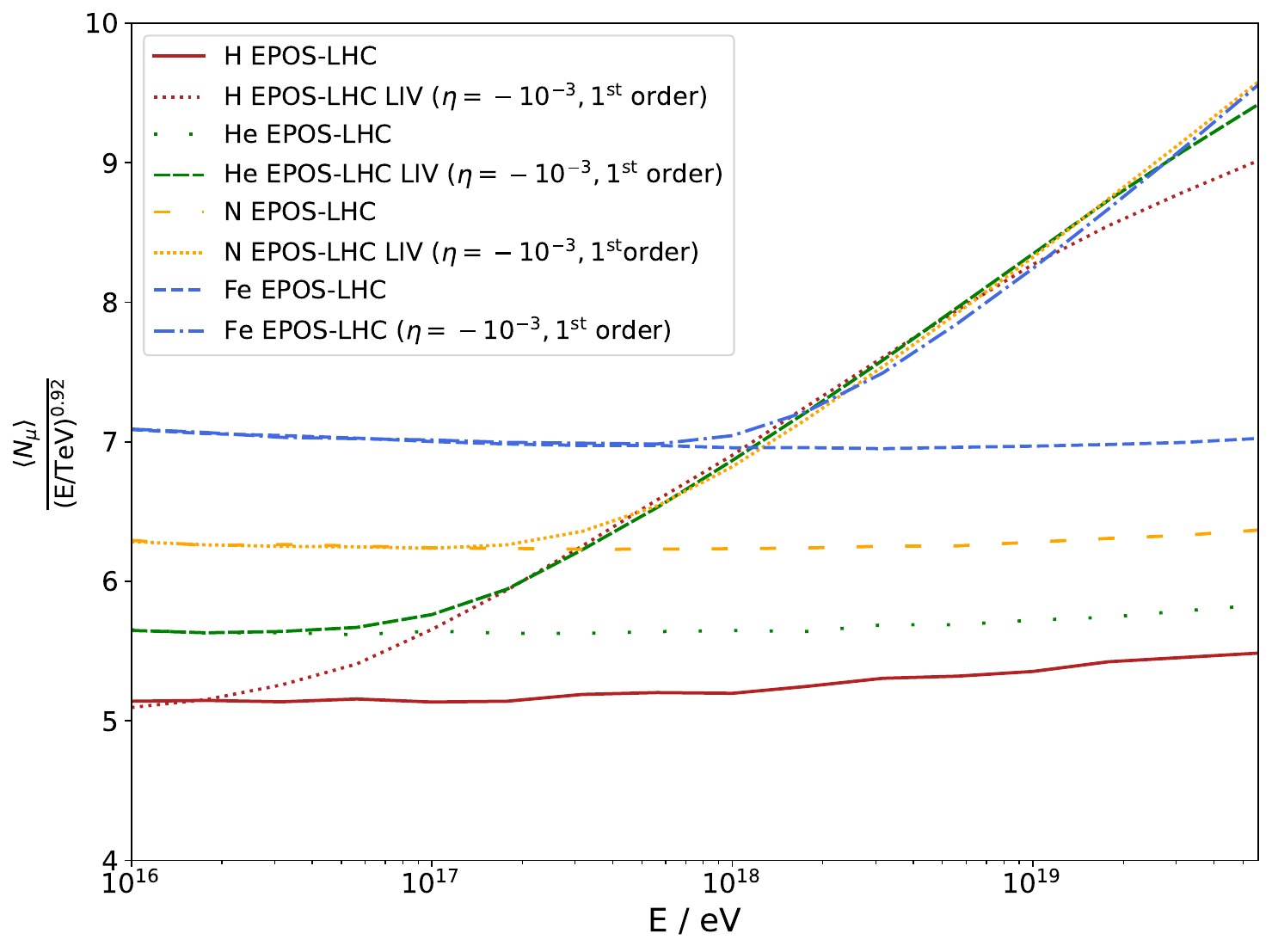}
 
 \caption{\label{fig:MuonAverage} Average number of muons at ground vs primary energy in LI and LIV cases.}
\end{figure}
To quantify the effect of LIV on the shower development, we have performed detailed simulations of its longitudinal profile by using the CONEX software~\cite{Pierog:2004re,Bergmann:2006yz} for the LI and LIV cases. For the LIV scenario, the software has been modified by changing the lifetime of any unstable particle according to Eq.~\ref{eq.gamma}. The  $\eta^{(n)}$  values considered in this study span the range from $-10^{-1}$ to $-10^{-6}$  in logarithmic steps, with order of violation  $n=1$.  For each value of $\eta$, 15000 primary cosmic-ray particles have been produced in the energy range between $10^{14}$~eV and $10^{21}$~eV, using EPOS-LHC~\cite{Pierog:2013ria} and QGSJetII-04~\cite{Ostapchenko:2010vb} hadronic interaction models and for different primary particle types i.e. hydrogen, helium, nitrogen and iron nuclei.

\begin{figure}[t]
 \centering
   \includegraphics[width=1.\columnwidth]{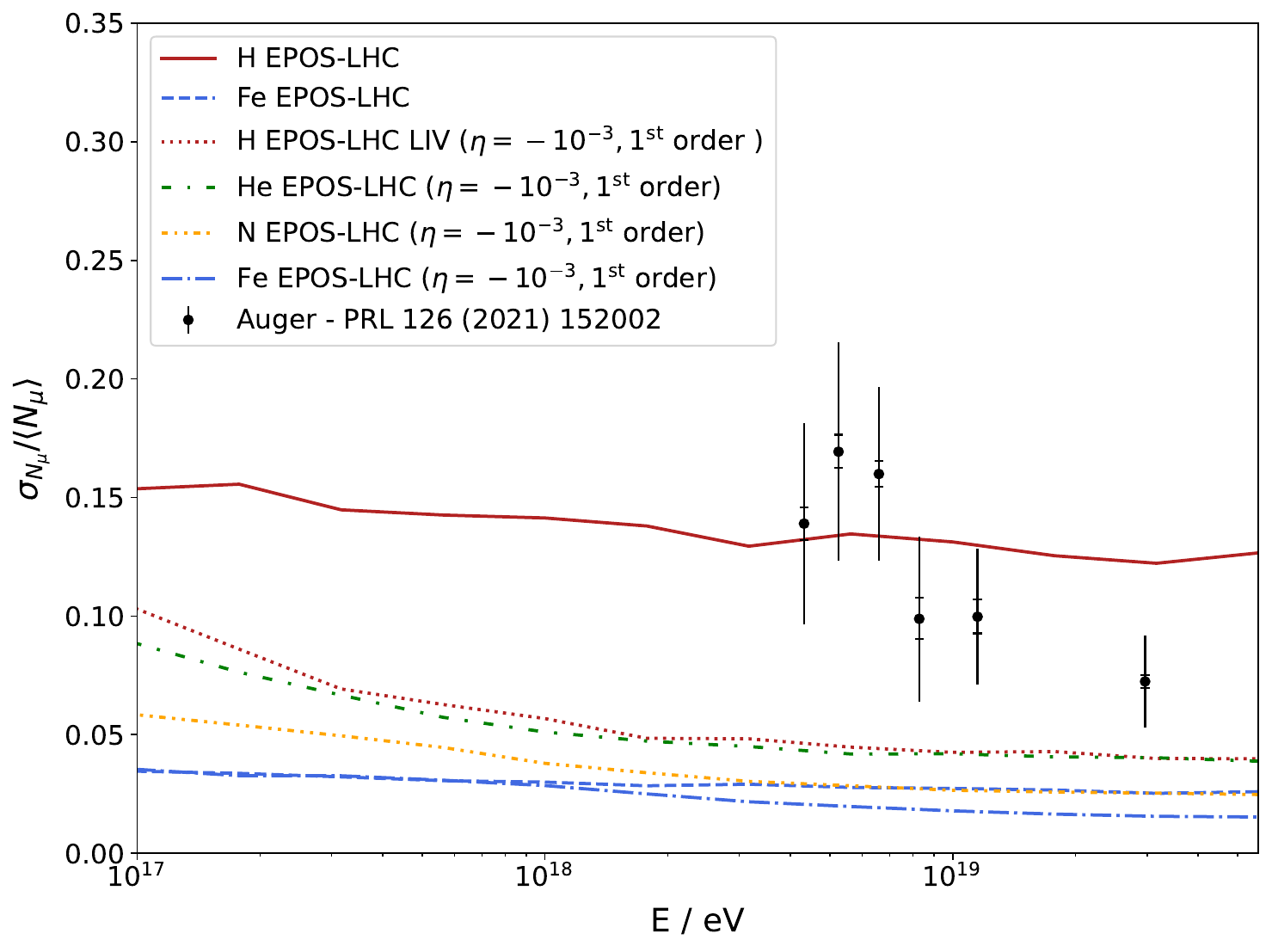}
 \caption{\label{fig:MuonFluctuation} Relative fluctuation of the number of muons vs primary energy; data (black points, from~\cite{PierreAuger:2021qsd}; the statistical uncertainty is indicated by the error bars, while the total systematic uncertainty by the square brackets) are compared to expectations from LI and LIV ($\eta=-10^{-3}$) models.}
\end{figure}

The enhanced number of interactions of neutral pions due to LIV contributes to the growth of the hadronic component in the cascade producing an increase in the number of muons $N_{\mu}$, as shown in Fig.~\ref{fig:MuonAverage}, where the average number of muons at ground as a function of the primary energy in LI and LIV cases are shown.  In the LIV scenarios considered here, the number of muons increases for any primary mass once the energy threshold is surpassed, where pions in the shower become stable. For iron primaries, this threshold occurs at higher energies due to their larger mass, as can be understood through a simple superposition model for nuclei with $A > 1$. Once LIV is in place, the rate of muon increase is similar for protons, iron, and intermediate masses. However, at the highest energies, the number of muons for heavier nuclei exceeds that of protons, as protons saturate earlier, when all available pions in the showers are interacting instead of decaying. This fact affects the fluctuations of the number of muons. In fact, the ratio of the fluctuations to the average number of muons (hereinafter referred to as relative fluctuations), dominated mostly by the first interaction~\cite{Cazon:2018gww}, considerably decreases in the presence of LIV for the case of protons, as shown in Fig.~\ref{fig:MuonFluctuation}. The decrease is due at the same time to the increase of the $N_{\mu}$ and to the decrease of fluctuations induced by LIV effects. The decrease is milder for heavier nuclear species.

Limits on the LIV parameter $\eta$ can be derived by comparing the expected decrease of the relative fluctuations caused by LIV effects with the experimental measurements~\cite{PierreAuger:2021qsd}.

\section*{\label{par:Results}Results and Discussion}

The relative fluctuations in the number of muons are directly linked to the mass composition of UHECRs, which consist of a mixture of nuclear species arriving at Earth~\cite{PierreAuger:2014sui,PierreAuger:2014gko}. 
An accurate determination of the contribution of different nuclear species to the UHECR flux is challenging due to uncertainties in the hadronic interaction models used to describe the shower development at these extreme energies. These models, which describe key processes such as particle production and energy dissipation, significantly influence the inferred composition. Thus, any attempt to constrain the degree of LIV in EAS development must account for these uncertainties.

To this aim, we notice that in the case of a composition comprised of the lightest and heaviest nuclear species (p and Fe respectively), the width of the distribution of the shower maxima is the largest, as shown for instance in~\cite{Kampert:2012mx}. The same behaviour can be expected for the relative muon fluctuations. Although when LIV is included the relative fluctuations in the number of muons differ only slightly among different primaries (as shown in the highest energy range in Fig.~\ref{fig:MuonFluctuation}), they are maximized when considering a mixed composition of protons and iron nuclei. 
The robustness of this choice against alternative mixtures, including intermediate masses, is illustrated in the Supplemental Material~\cite{supplemental_material} through the corresponding so-called umbrella plots which illustrate the allowed phase-space of fluctuations average and average of the number of muons. These show that, in the region relevant to this analysis, the hierarchy of relative muon fluctuations is largely preserved and intermediate-mass combinations do not exceed the proton-iron benchmark. This makes the proton-iron mixture a conservative benchmark for the present exclusion procedure.

As described in~\cite{PierreAuger:2021qsd}, the average number of muons, $\langle N_{\mu}\rangle_{\text{mix}}$, and its fluctuations, $\mathrm{\mathrm{\sigma}}_{N_{\mu},\text{mix}}$, can be calculated  for a mixture of p and Fe  as follows:
\begin{equation}\begin{split}
\langle N_{\mu}\rangle_{\text{mix}}(f;\eta)& =(1-f)\langle N_{\mu}\rangle_{\mathrm{p}} + f \langle N_{\mu}\rangle_{\mathrm{Fe}}\\
\mathrm{\sigma}^2_{N_{\mu},\text{mix}}(f;\eta) &= (1-f)\mathrm{\sigma}^2_{N_{\mu},\mathrm{p}}+\\ f \mathrm{\sigma}^2_{N_{\mu},\mathrm{Fe}} & + f(1-f)(\langle N_{\mu}\rangle_{\mathrm{p}} - \langle N_{\mu}\rangle _{\mathrm{Fe}})^2
\label{eq:mixedComp}
\end{split}\end{equation} 
where $f$, which depends on the energy, is the relative abundance of iron nuclei~\cite{footnote6}. The average number of muons $\langle N_{\mu}\rangle$ and the $\mathrm{\sigma}^2_{N_{\mu}}$ for air showers induced by both proton and iron nuclei have been obtained from the air shower simulations performed with different LIV parameters over the energy range reported in the previous section. These results have been used to parametrize $\langle N_{\mu}\rangle$ and  $\mathrm{\sigma}_{N_{\mu}}$ across the range $E=10^{14}$ eV to $E=10^{20}$ eV, for $-10^{-1}<\eta<-10^{-6}$ and for both proton and iron primaries. These parametrisations were then extrapolated to provide expectations in the energy range  $10^{16}$  to  $10^{20}$~eV and for  $\eta$  values extended down to  $-10^{-16}$  (see Appendix). For any given value of  $\eta$, the parametrisations allow us to calculate the expected relative fluctuations for a specific mixture of proton and iron nuclei (hereinafter referred to as mixed relative fluctuations) as follows:

\begin{equation}
\frac{\sigma_{N_\mu}}{\langle N_{\mu}\rangle}(f;\eta)=\frac{\sqrt{\mathrm{\sigma}^2_{{N_{\mu}},\text{mix}}(N_{\mu})(f;\eta)}} {\langle N_{\mu}\rangle_{\text{mix}}(f;\eta)}.
\label{eq:RelativeFluct}
\end{equation}
To identify the most conservative LIV model, the relative fluctuations are analyzed as a function of energy by determining the value of  $f$ that maximizes the mixed fluctuations for each energy bin. This maximum, determined by varying  $f$  between 0 and 1 (i.e., scanning from a pure proton to a pure iron composition), is found to be consistent across all tested values of the LIV parameter  $\eta$ (see Appendix). Consequently, the most conservative LIV model for any composition corresponds to the values $f(\text{E})$ that maximise Eq.~\ref{eq:RelativeFluct}, under the constraint that the predicted curves lie entirely below the measured data. Notably, this procedure allows the most conservative LIV relative fluctuations to be determined for any given LIV parameter value without requiring additional shower simulations. 

\begin{figure}[!t]
 \centering
    \includegraphics[width=1.\columnwidth]{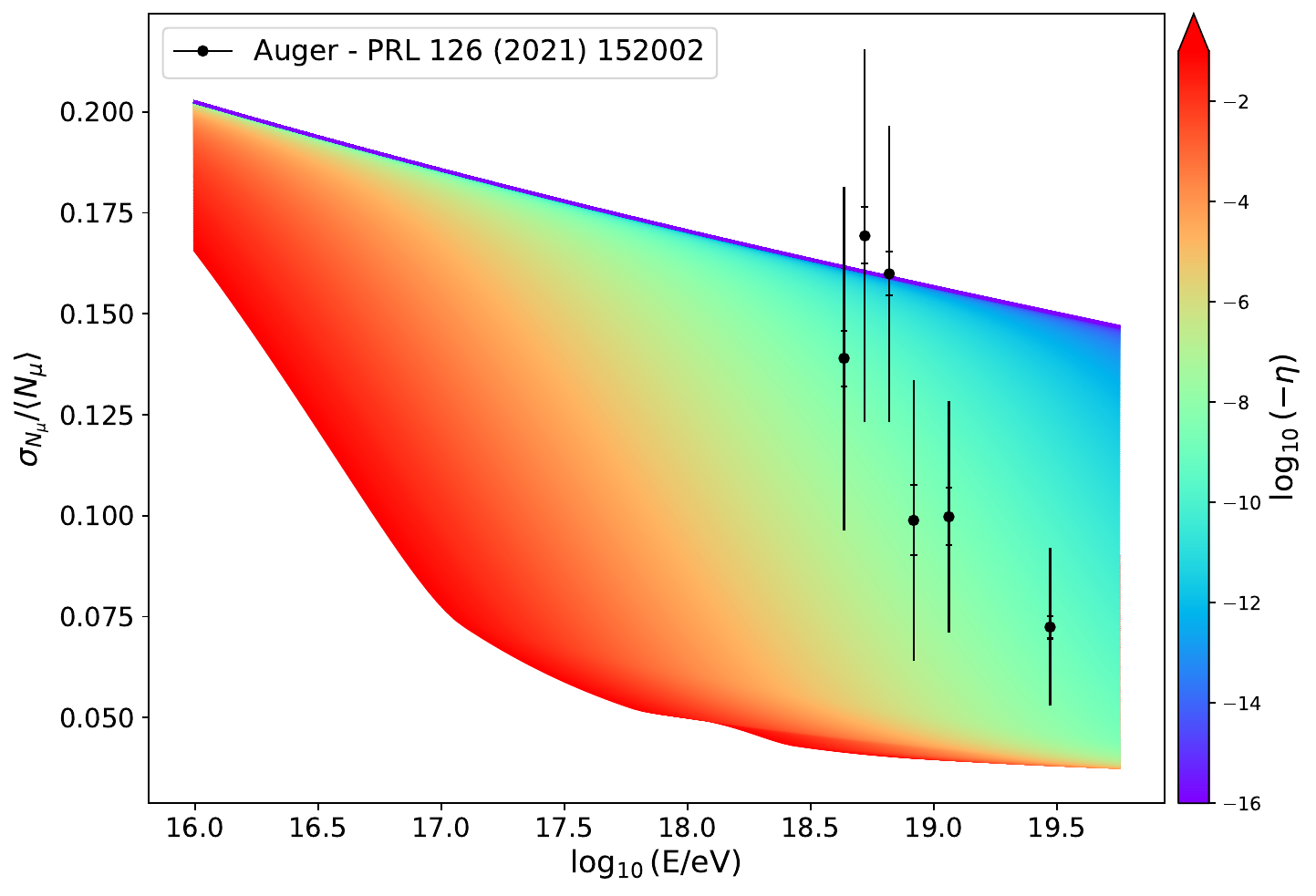}
 \caption{\label{fig:RelativeFluctuationVsEnergy} Mixed relative fluctuations obtained using the parameterisations considering $\eta$ in the range $[-10^{-1},-10^{-16}]$ as a function of the primary energy, corresponding to the maximum with respect to $f$. Each colour corresponds to a different violation strength (right axis, where the arrow indicates the direction of increasing violation strength). The measured relative fluctuations in the number of muons are shown with black points~\cite{PierreAuger:2021qsd}.}
\end{figure}

In Fig.~\ref{fig:RelativeFluctuationVsEnergy}, the colour code represents the relative fluctuations corresponding to the maxima with respect to $f$, derived from the parametrisations considering $\eta$ values in the range [$-10^{-1}$, $-10^{-16}$]. By selecting only the mixed relative fluctuations below the experimental data points, the $\chi^2$ is computed as a function of  $\eta$. This approach yields a continuous confidence level for excluding specific LIV models, allowing the determination of the strictest lower bound on $\eta$.

The mixed relative fluctuations have been estimated for three different confidence levels (CLs) by considering all available experimental data. The exclusion limits for $\eta$, quantified as $\log_{10}(-\eta)$, represent the smallest level of violation that can be ruled out at each corresponding confidence level. The exclusion limits at the 99.9\%, 95.5\% and 90.5\% CLs  are summarized in Table~\ref{tab:limit_label}.
These bounds were determined independently using two distinct hadronic interaction models, EPOS-LHC and QGSJetII-04. The quoted central value is the mean of the two results, while the spread between them is taken as an estimate of the systematic uncertainty associated with the hadronic interaction modeling.

It can be observed that, if the discrepancy in the energy reconstruction between the LIV scenario and the standard one were explicitly accounted for, a shift in the experimental data towards higher reconstructed energies would occur. This is because, in the LIV scenario, a smaller fraction of the primary energy is transferred to the electromagnetic component of the shower, which is used to reconstruct the energy. However, this bias in the energy estimation between the two scenarios remains below 5\% for all the considered $\eta$ parameter values. As a result, incorporating this effect into the analysis would change the reconstructed energy values leading to a more stringent constraint on the $\eta$ parameter.

\begin{table}[t]
    \centering
    \renewcommand{\arraystretch}{1.4}
    \setlength{\tabcolsep}{8pt} 
    \begin{tabular}{c|c|c|c}
        \hline
        \textbf{C.L.} & \textbf{90.5\%} & \textbf{95.5\%} & \textbf{99.9\%} \\ \hline
        $\log_{10}(-\eta)$ & $-7.03^{+0.21}_{-0.50}$ & $-6.82^{+0.19}_{-0.54}$ & $-5.90^{+0.13}_{-0.62}$ \\ 
        \hline
    \end{tabular}
    \caption{Lower limits on $\log_{10}(-\eta)$ derived from this analysis at different confidence levels (CL). The asymmetric uncertainties indicate the associated systematic uncerainty from the spread between the EPOS-LHC and QGSJetII-04 predictions.}
    \label{tab:limit_label}
\end{table}

\section*{\label{par:conclus}Conclusions}
In this work, interactions of particles at the most extreme energies have been exploited to test Lorentz invariance, a frontier in physics which is reminiscent of the early 20th century, when cosmic rays paved the future of particle physics. 
A significant milestone in constraining first-order Lorentz invariance violation has been achieved through the analysis of muon fluctuations within extensive air showers - used for the first time to this purpose - measured at the Pierre Auger Observatory. Using a conservative model based on mixed nuclear composition of protons and iron nuclei, we derived the most stringent LIV bounds, for negative values of the LIV parameter: $\eta_{\pi}>-1.26\times10^{-6}$ at 5$\sigma$ C.L. Upper bounds are instead obtained with studies considering UHECR propagation, which find that $\eta_{\pi}<2.4\times10^{-10}$ at 5$\sigma$ C.L. In addition to showing the potential of muon-related observables to detect deviations from Lorentz invariance, with this work we demonstrate the unique capability of testing both negative and positive values of LIV parameters exploiting at most the data of the same observatory. Future efforts will be devoted to extend this analysis to investigating second-order LIV effects and combining muon fluctuation data with mass composition estimates and other observables, such as radio detector data~\cite{Salamida:2023} and underground muon detector (UMD) measurements~\cite{Salamida:2023}. 

In summary, this work represents a foundational step towards a comprehensive test of Lorentz invariance at ultra-high energies, as aimed in~\cite{AlvesBatista:2023wqm}, thanks to the unique capability of the Pierre Auger Observatory to explore extreme energy scales.
\section*{Acknowledgments}

\begin{sloppypar}
The successful installation, commissioning, and operation of the Pierre Auger Observatory would not have been possible without the strong commitment and effort from the technical and administrative staff in Malarg\"ue. We are very grateful to the following agencies and organizations for financial support:
\end{sloppypar}

\begin{sloppypar}
Argentina -- Comisi\'on Nacional de Energ\'\i{}a At\'omica; Agencia Nacional de
Promoci\'on Cient\'\i{}fica y Tecnol\'ogica (ANPCyT); Consejo Nacional de
Investigaciones Cient\'\i{}ficas y T\'ecnicas (CONICET); Gobierno de la
Provincia de Mendoza; Municipalidad de Malarg\"ue; NDM Holdings and Valle
Las Le\~nas; in gratitude for their continuing cooperation over land
access; Australia -- the Australian Research Council; Belgium -- Fonds
de la Recherche Scientifique (FNRS); Research Foundation Flanders (FWO),
Marie Curie Action of the European Union Grant No.~101107047; Brazil --
Minist\'erio da Ci\^encia, Tecnologia e Inova\c{c}\~ao (MCTI); Czech Republic --
GACR 24-13049S, CAS LQ100102401, MEYS LM2023032,
CZ.02.1.01/0.0/0.0/16{\textunderscore}013/0001402, CZ.02.1.01/0.0/0.0/18{\textunderscore}046/0016010
and CZ.02.1.01/0.0/0.0/17{\textunderscore}049/0008422 and
CZ.02.01.01/00/22{\textunderscore}008/0004632; France -- Centre de Calcul IN2P3/CNRS;
Centre National de la Recherche Scientifique (CNRS); Institut National
de Physique Nucl\'eaire et de Physique des Particules (IN2P3/CNRS);
Germany -- Bundesministerium f\"ur Forschung, Technologie und Raumfahrt
(BMFTR); Deutsche Forschungsgemeinschaft (DFG); Ministerium f\"ur Finanzen
Baden-W\"urttemberg; Helmholtz Alliance for Astroparticle Physics (HAP);
Hermann von Helmholtz-Gemeinschaft Deutscher Forschungszentren e.V.;
Ministerium f\"ur Kultur und Wissenschaft des Landes Nordrhein-Westfalen;
Ministerium f\"ur Wissenschaft, Forschung und Kunst des Landes
Baden-W\"urttemberg; Italy -- Istituto Nazionale di Fisica Nucleare
(INFN); Istituto Nazionale di Astrofisica (INAF); Ministero
dell'Universit\`a e della Ricerca (MUR); CETEMPS Center of Excellence;
Ministero degli Affari Esteri (MAE), ICSC Centro Nazionale di Ricerca in
High Performance Computing, Big Data and Quantum Computing, funded by
European Union NextGenerationEU, reference code CN{\textunderscore}00000013; M\'exico --
Consejo Nacional de Ciencia y Tecnolog\'\i{}a (CONACYT-SECHTI)
No.~CB-A1-S-46703, Universidad Nacional Aut\'onoma de M\'exico (UNAM)
PAPIIT-IN114924; Benem\'erita Universidad Aut\'onoma de Puebla (BUAP), VIEP
and Laboratorio Nacional de Superc\'omputo del Sureste de M\'exico (LNS);
and Benem\'erita Universidad Aut\'onoma de Chiapas (UNACH); The Netherlands
-- Ministry of Education, Culture and Science; Netherlands Organisation
for Scientific Research (NWO); Dutch national e-infrastructure with the
support of SURF Cooperative; Poland -- Ministry of Science and Higher
Education, grant No.~2022/WK/12; National Science Centre, grants
No.~2020/39/B/ST9/01398, and 2022/45/B/ST9/02163; Portugal -- Portuguese
national funds and FEDER funds within Programa Operacional Factores de
Competitividade through Funda\c{c}\~ao para a Ci\^encia e a Tecnologia
(COMPETE); Romania -- Ministry of Education and Research, contract
no.~30N/2023 under Romanian National Core Program LAPLAS VII, and grant
no.~PN 23 21 01 02; Slovenia -- Slovenian Research and Innovation
Agency, grants P1-0031, I0-0033; Spain -- Ministerio de Ciencia,
Innovaci\'on y Universidades/Agencia Estatal de Investigaci\'on MICIU/AEI
/10.13039/501100011033 (PID2022-140510NB-I00, PCI2023-145952-2,
CNS2024-154676, and Mar\'\i{}a de Maeztu CEX2023-001318-M), Xunta de Galicia
(CIGUS Network of Research Centers, Consolidaci\'on ED431C-2025/11 and
ED431F-2022/15) and European Union ERDF; USA -- Department of Energy,
Contracts No.~DE-AC02-07CH11359, No.~DE-FR02-04ER41300,
No.~DE-FG02-99ER41107 and No.~DE-SC0011689; National Science Foundation,
Grant No.~0450696, and NSF-2013199; The Grainger Foundation;
Astrophysics Centre for Multi-messenger studies in Europe (ACME) EU
Grant No 101131928; and UNESCO.
\end{sloppypar}

\bibliography{biblio}
\newpage
%%%%%%%%%%%%%%%%%%%%%%%%%%%%%%%%%%%%%%%%%%%%%%%%%%%%%%%%%%%%%%%%%%%%%%%%%%%%%%%%%%%%%%
% AUTHOR-LIST
\onecolumngrid

\begin{center}
\rule{0.1\columnwidth}{0.5pt}
\raisebox{-0.4ex}{\scriptsize$\bullet$}
\rule{0.1\columnwidth}{0.5pt}
\end{center}
\vspace{3mm}

\begin{wrapfigure}[8]{l}{0.06\textwidth}
\vspace{-2.9ex}
\includegraphics[width=1.4\linewidth]{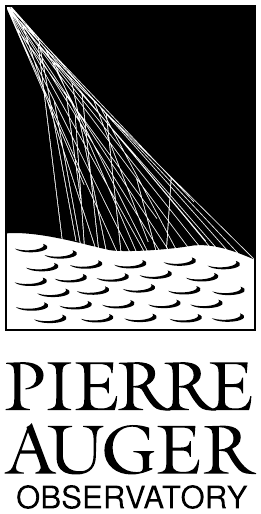}
\end{wrapfigure}
\begin{sloppypar}\noindent
% created on 2026-03-13
A.~Abdul Halim$^{13}$,
P.~Abreu$^{67}$,
M.~Aglietta$^{50,49}$,
I.~Allekotte$^{1}$,
K.~Almeida Cheminant$^{75,74}$,
R.~Aloisio$^{42,43}$,
J.~Alvarez-Mu\~niz$^{73}$,
A.~Ambrosone$^{42}$,
J.~Ammerman Yebra$^{73}$,
L.~Anchordoqui$^{79}$
% ...

\end{sloppypar}
\begin{center}
\par\noindent
\textbf{The Pierre Auger Collaboration}
\end{center}

\vspace{1ex}
% created on 2026-03-13
% needs \usepackage{enumitem}
\begin{description}[labelsep=0.2em,align=right,labelwidth=0.7em,labelindent=0em,leftmargin=2em,noitemsep,before={\renewcommand\makelabel[1]{##1 }}]
\item[$^{1}$] Centro At\'omico Bariloche and Instituto Balseiro (CNEA-UNCuyo-CONICET), San Carlos de Bariloche, Argentina
\item[$^{2}$] Departamento de F\'\i{}sica and Departamento de Ciencias de la Atm\'osfera y los Oc\'eanos, FCEyN, Universidad de Buenos Aires and CONICET, Buenos Aires, Argentina
\item[$^{3}$] IFLP, Universidad Nacional de La Plata and CONICET, La Plata, Argentina
\item[$^{4}$] Instituto de Astronom\'\i{}a y F\'\i{}sica del Espacio (IAFE, CONICET-UBA), Buenos Aires, Argentina
\item[$^{5}$] Instituto de F\'\i{}sica de Rosario (IFIR) -- CONICET/U.N.R.\ and Facultad de Ciencias Bioqu\'\i{}micas y Farmac\'euticas U.N.R., Rosario, Argentina
\item[$^{6}$] Instituto de Tecnolog\'\i{}as en Detecci\'on y Astropart\'\i{}culas (CNEA, CONICET, UNSAM), and Universidad Tecnol\'ogica Nacional -- Facultad Regional Mendoza (CONICET/CNEA), Mendoza, Argentina
\item[$^{7}$] Instituto de Tecnolog\'\i{}as en Detecci\'on y Astropart\'\i{}culas (CNEA, CONICET, UNSAM), Buenos Aires, Argentina
\item[$^{8}$] International Center of Advanced Studies and Instituto de Ciencias F\'\i{}sicas, ECyT-UNSAM and CONICET, Campus Miguelete -- San Mart\'\i{}n, Buenos Aires, Argentina
\item[$^{9}$] Laboratorio Atm\'osfera -- Departamento de Investigaciones en L\'aseres y sus Aplicaciones -- UNIDEF (CITEDEF-CONICET), Argentina
\item[$^{10}$] Observatorio Pierre Auger, Malarg\"ue, Argentina
\item[$^{11}$] Observatorio Pierre Auger and Comisi\'on Nacional de Energ\'\i{}a At\'omica, Malarg\"ue, Argentina
\item[$^{12}$] Universidad Tecnol\'ogica Nacional -- Facultad Regional Buenos Aires, Buenos Aires, Argentina
\item[$^{13}$] Adelaide University, Adelaide, S.A., Australia
\item[$^{14}$] Universit\'e Libre de Bruxelles (ULB), Brussels, Belgium
\item[$^{15}$] Vrije Universiteit Brussels, Brussels, Belgium
\item[$^{16}$] Centro Brasileiro de Pesquisas Fisicas, Rio de Janeiro, RJ, Brazil
\item[$^{17}$] Centro Federal de Educa\c{c}\~ao Tecnol\'ogica Celso Suckow da Fonseca, Petropolis, Brazil
\item[$^{18}$] Universidade de S\~ao Paulo, Escola de Engenharia de Lorena, Lorena, SP, Brazil
\item[$^{19}$] Universidade de S\~ao Paulo, Instituto de F\'\i{}sica de S\~ao Carlos, S\~ao Carlos, SP, Brazil
\item[$^{20}$] Universidade de S\~ao Paulo, Instituto de F\'\i{}sica, S\~ao Paulo, SP, Brazil
\item[$^{21}$] Universidade Estadual de Campinas (UNICAMP), IFGW, Campinas, SP, Brazil
\item[$^{22}$] Universidade Estadual de Feira de Santana, Feira de Santana, Brazil
\item[$^{23}$] Universidade Federal do ABC, Santo Andr\'e, SP, Brazil
\item[$^{24}$] Universidade Federal do Paran\'a, Setor Palotina, Palotina, Brazil
\item[$^{25}$] Universidade Federal do Rio de Janeiro, Instituto de F\'\i{}sica, Rio de Janeiro, RJ, Brazil
\item[$^{26}$] Universidad de Medell\'\i{}n, Medell\'\i{}n, Colombia
\item[$^{27}$] Universidad Industrial de Santander, Bucaramanga, Colombia
\item[$^{28}$] Charles University, Faculty of Mathematics and Physics, Institute of Particle and Nuclear Physics, Prague, Czech Republic
\item[$^{29}$] Institute of Physics of the Czech Academy of Sciences, Prague, Czech Republic
\item[$^{30}$] Palacky University, Olomouc, Czech Republic
\item[$^{31}$] CNRS/IN2P3, IJCLab, Universit\'e Paris-Saclay, Orsay, France
\item[$^{32}$] Laboratoire de Physique Nucl\'eaire et de Hautes Energies (LPNHE), Sorbonne Universit\'e, Universit\'e de Paris, CNRS-IN2P3, Paris, France
\item[$^{33}$] Universit\'e Paris-Saclay, CNRS/IN2P3, IJCLab, Orsay, France
\item[$^{34}$] Bergische Universit\"at Wuppertal, Department of Physics, Wuppertal, Germany
\item[$^{35}$] Karlsruhe Institute of Technology (KIT), Institute for Experimental Particle Physics, Karlsruhe, Germany
\item[$^{36}$] Karlsruhe Institute of Technology (KIT), Institut f\"ur Prozessdatenverarbeitung und Elektronik, Karlsruhe, Germany
\item[$^{37}$] Karlsruhe Institute of Technology (KIT), Institute for Astroparticle Physics, Karlsruhe, Germany
\item[$^{38}$] RWTH Aachen University, III.\ Physikalisches Institut A, Aachen, Germany
\item[$^{39}$] TU Dortmund University, Department of Physics, Dortmund, Germany
\item[$^{40}$] Universit\"at Hamburg, II.\ Institut f\"ur Theoretische Physik, Hamburg, Germany
\item[$^{41}$] Universit\"at Siegen, Department Physik -- Experimentelle Teilchenphysik, Siegen, Germany
\item[$^{42}$] Gran Sasso Science Institute, L'Aquila, Italy
\item[$^{43}$] INFN Laboratori Nazionali del Gran Sasso, Assergi (L'Aquila), Italy
\item[$^{44}$] INFN, Sezione di Catania, Catania, Italy
\item[$^{45}$] INFN, Sezione di Lecce, Lecce, Italy
\item[$^{46}$] INFN, Sezione di Milano, Milano, Italy
\item[$^{47}$] INFN, Sezione di Napoli, Napoli, Italy
\item[$^{48}$] INFN, Sezione di Roma ``Tor Vergata'', Roma, Italy
\item[$^{49}$] INFN, Sezione di Torino, Torino, Italy
\item[$^{50}$] Osservatorio Astrofisico di Torino (INAF), Torino, Italy
\item[$^{51}$] Politecnico di Milano, Dipartimento di Scienze e Tecnologie Aerospaziali , Milano, Italy
\item[$^{52}$] Universit\`a del Salento, Dipartimento di Matematica e Fisica ``E.\ De Giorgi'', Lecce, Italy
\item[$^{53}$] Universit\`a dell'Aquila, Dipartimento di Scienze Fisiche e Chimiche, L'Aquila, Italy
\item[$^{54}$] Universit\`a di Catania, Dipartimento di Fisica e Astronomia ``Ettore Majorana``, Catania, Italy
\item[$^{55}$] Universit\`a di Milano, Dipartimento di Fisica, Milano, Italy
\item[$^{56}$] Universit\`a di Napoli ``Federico II'', Dipartimento di Fisica ``Ettore Pancini'', Napoli, Italy
\item[$^{57}$] Universit\`a di Palermo, Dipartimento di Fisica e Chimica ''E.\ Segr\`e'', Palermo, Italy
\item[$^{58}$] Universit\`a di Roma ``Tor Vergata'', Dipartimento di Fisica, Roma, Italy
\item[$^{59}$] Universit\`a Torino, Dipartimento di Fisica, Torino, Italy
\item[$^{60}$] Benem\'erita Universidad Aut\'onoma de Puebla, Puebla, M\'exico
\item[$^{61}$] Unidad Profesional Interdisciplinaria en Ingenier\'\i{}a y Tecnolog\'\i{}as Avanzadas del Instituto Polit\'ecnico Nacional (UPIITA-IPN), M\'exico, D.F., M\'exico
\item[$^{62}$] Universidad Aut\'onoma de Chiapas, Tuxtla Guti\'errez, Chiapas, M\'exico
\item[$^{63}$] Universidad Michoacana de San Nicol\'as de Hidalgo, Morelia, Michoac\'an, M\'exico
\item[$^{64}$] Universidad Nacional Aut\'onoma de M\'exico, M\'exico, D.F., M\'exico
\item[$^{65}$] Institute of Nuclear Physics PAN, Krakow, Poland
\item[$^{66}$] University of \L{}\'od\'z, Faculty of High-Energy Astrophysics,\L{}\'od\'z, Poland
\item[$^{67}$] Laborat\'orio de Instrumenta\c{c}\~ao e F\'\i{}sica Experimental de Part\'\i{}culas -- LIP and Instituto Superior T\'ecnico -- IST, Universidade de Lisboa -- UL, Lisboa, Portugal
\item[$^{68}$] ``Horia Hulubei'' National Institute for Physics and Nuclear Engineering, Bucharest-Magurele, Romania
\item[$^{69}$] Institute of Space Science, Bucharest-Magurele, Romania
\item[$^{70}$] Center for Astrophysics and Cosmology (CAC), University of Nova Gorica, Nova Gorica, Slovenia
\item[$^{71}$] Experimental Particle Physics Department, J.\ Stefan Institute, Ljubljana, Slovenia
\item[$^{72}$] Universidad de Granada and C.A.F.P.E., Granada, Spain
\item[$^{73}$] Instituto Galego de F\'\i{}sica de Altas Enerx\'\i{}as (IGFAE), Universidade de Santiago de Compostela, Santiago de Compostela, Spain
\item[$^{74}$] IMAPP, Radboud University Nijmegen, Nijmegen, The Netherlands
\item[$^{75}$] Nationaal Instituut voor Kernfysica en Hoge Energie Fysica (NIKHEF), Science Park, Amsterdam, The Netherlands
\item[$^{76}$] Stichting Astronomisch Onderzoek in Nederland (ASTRON), Dwingeloo, The Netherlands
\item[$^{77}$] Case Western Reserve University, Cleveland, OH, USA
\item[$^{78}$] Colorado School of Mines, Golden, CO, USA
\item[$^{79}$] Department of Physics and Astronomy, Lehman College, City University of New York, Bronx, NY, USA
\item[$^{80}$] Michigan Technological University, Houghton, MI, USA
\item[$^{81}$] New York University, New York, NY, USA
\item[$^{82}$] University of Chicago, Enrico Fermi Institute, Chicago, IL, USA
\item[$^{83}$] University of Delaware, Department of Physics and Astronomy, Bartol Research Institute, Newark, DE, USA
\item[] -----
\item[$^{a}$] Max-Planck-Institut f\"ur Radioastronomie, Bonn, Germany
\item[$^{b}$] also at Kapteyn Institute, University of Groningen, Groningen, The Netherlands
\item[$^{c}$] School of Physics and Astronomy, University of Leeds, Leeds, United Kingdom
\item[$^{d}$] Fermi National Accelerator Laboratory, Fermilab, Batavia, IL, USA (Affiliation for identification purposes only)
\item[$^{e}$] Pennsylvania State University, University Park, PA, USA
\item[$^{f}$] Colorado State University, Fort Collins, CO, USA
\item[$^{g}$] Louisiana State University, Baton Rouge, LA, USA
\item[$^{h}$] Universit\'e Grenoble Alpes, CNRS, Grenoble Institute of Engineering, LPSC-IN2P3, Grenoble, France
\item[$^{i}$] now at Graduate School of Science, Osaka Metropolitan University, Osaka, Japan
\item[$^{j}$] Institut universitaire de France (IUF), France
\end{description}

%%%%%%%%%%%%%%%%%%%%%%%%%%%%%%%%%%%%%%%%%%%%%%%%%%%%%%%%%%%%%%%%%%%%%%%%%%%%%%%%%%%%%%

\cleardoublepage

\appendix

\twocolumngrid
\newpage
\onecolumngrid

\section{SUPPLEMENTAL MATERIAL: Bounds on Lorentz Invariance Violation from muon fluctuations at the Pierre Auger Observatory}

\section{I.~LIV effect on shower maximum and energy}
\label{sec:xmax}

\begin{figure}[h!]
    \centering
    \begin{minipage}{0.48\textwidth}
        \centering
        \includegraphics[width=\linewidth]{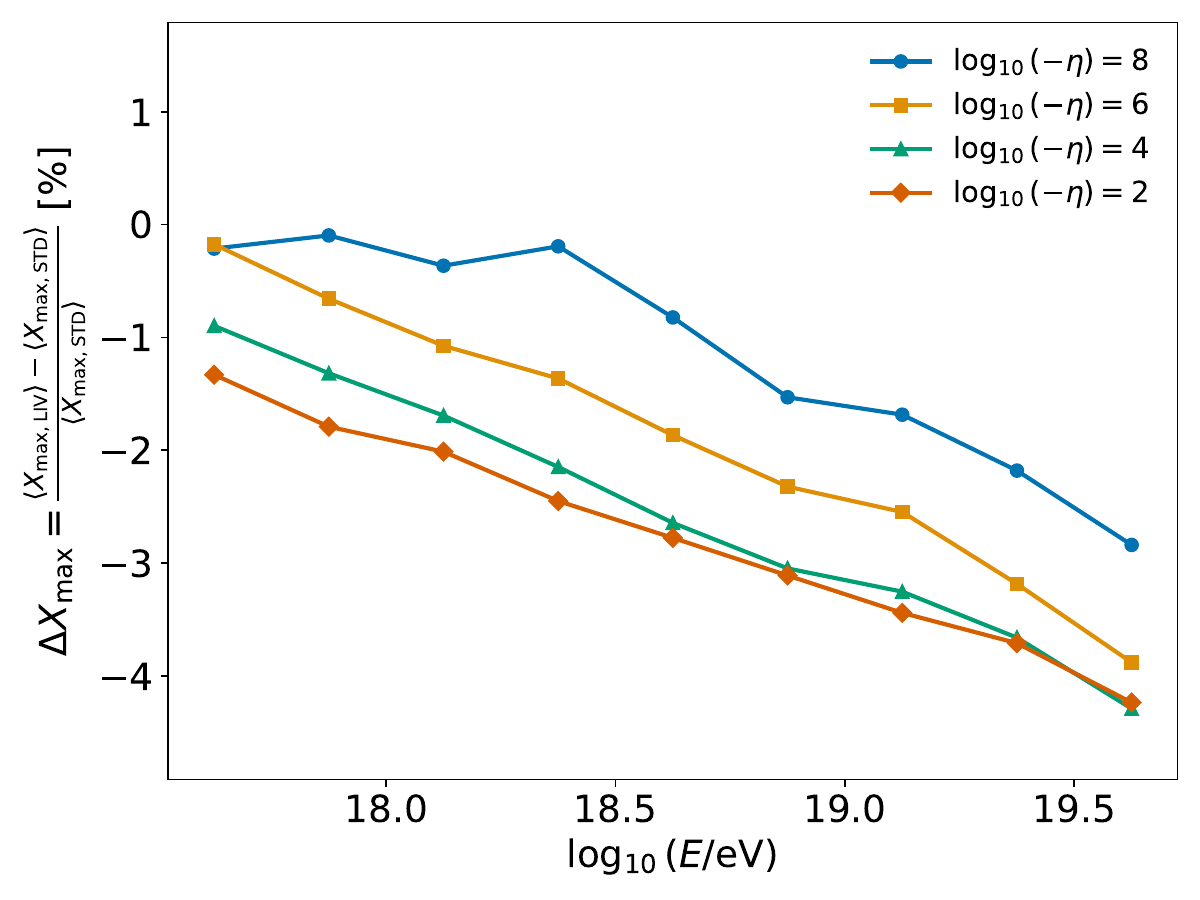}
    \end{minipage}
    \hfill
    \begin{minipage}{0.48\textwidth}
        \centering
        \includegraphics[width=\linewidth]{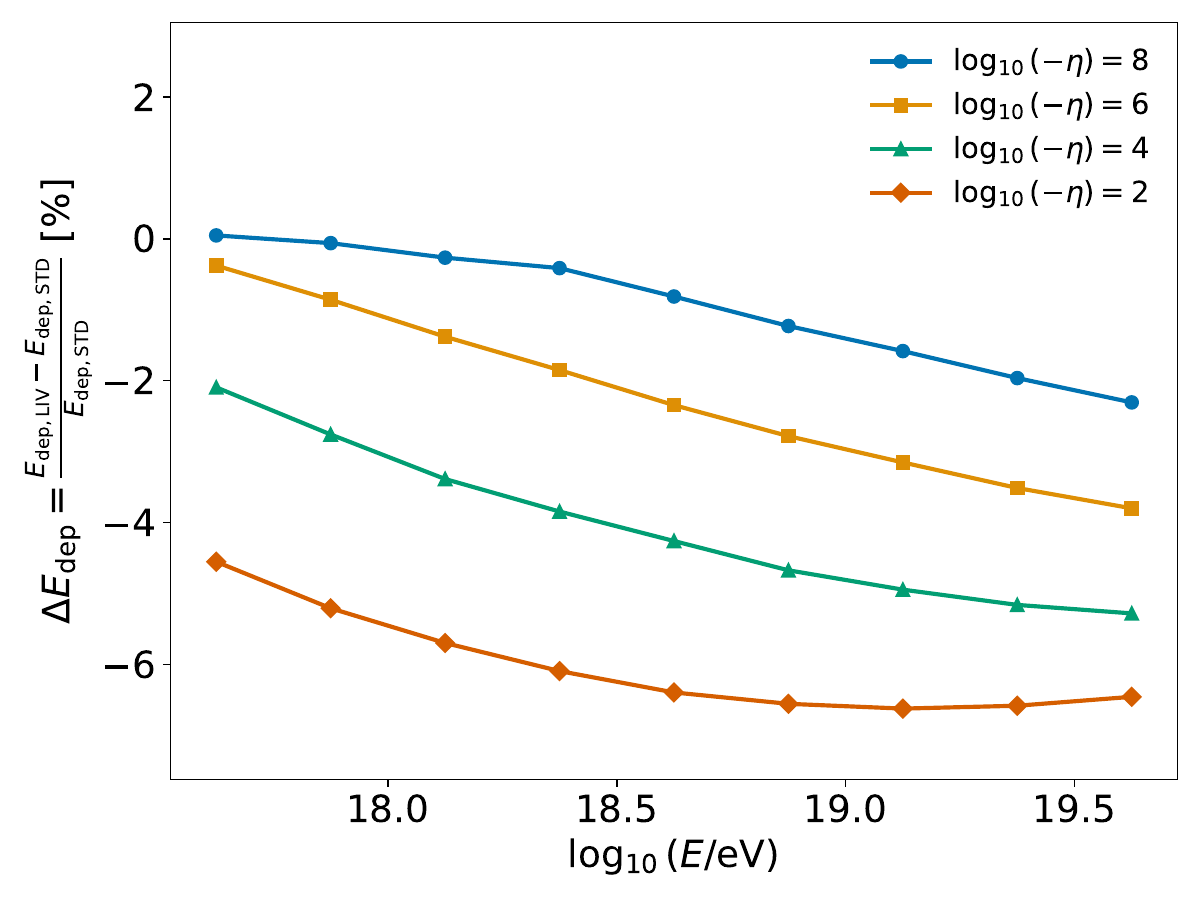}
    \end{minipage}
    \caption{Relative differences in shower maximum (left panel) and deposited energy (right panel) as a function of $\log_{10}(E/\mathrm{eV})$ for different LIV values, using the EPOS-LHC hadronic interaction model. Both are expressed in percent.}
    \label{fig:delta_combined}
\end{figure}

The depth of the shower maximum and the energy deposit in the atmosphere have been evaluated in the same set of CONEX simulations used for the muon analysis. In the presence of LIV, $\langle X_{\mathrm{max}}\rangle$ shows a systematic shift towards lower values with respect to the Lorentz-invariant case, particularly for proton primaries at the highest energies. However, the magnitude of this variation remains small compared to the changes observed in the muon sector. For the values of LIV $\eta$ parameter within the region of interest considered in this work, the shift towards shallower $\langle X_{\mathrm{max}}\rangle$ with respect to the standard case does not exceed 5$\%$. This behaviour is illustrated, as an example for EPOS-LHC hadronic interaction model, in Fig.~\ref{fig:delta_combined} (left panel).

A similar trend is found for the energy deposited in the atmosphere: the suppression of neutral-pion decay reduces the fraction of energy transferred to the electromagnetic component, leading to a modest reduction in the normalization of the longitudinal profile. Across the full range of constrained LIV parameters, the decrease in the deposited energy is consistently below 5$\%$, as shown in Fig.~\ref{fig:delta_combined} (right panel).

These results confirm that while the shower maximum and the deposited energy carry some sensitivity to LIV effects, their relative modifications are much less pronounced than for the muon content at ground level. Consequently, they provide complementary but less stringent observables in assessing possible deviations from Lorentz invariance.

\section{II.~Parameterisations of Muon Observables in LIV Models}
\label{sec:param}
The relative muon fluctuations, defined as $\sigma_{N_\mu}/\langle N_\mu \rangle$, are sensitive to energy, primary mass and hadronic interaction model. Since the necessary high statistics simulations are feasible only for a discrete set of energies and mass numbers, a functional representation is essential to enable a smooth evaluation of muon observables under LIV hypotheses. 
However, the parameterisations have been produced only for protons and iron nuclei which allow us to reproduce the most conservative LIV scenario, namely, the composition that yields the largest possible value of $\sigma_{N_\mu}/\langle N_\mu \rangle$ under the hypothesis of a given LIV strength. In the following, we describe the two-step fitting procedure that enables this construction and the parametrisation which are valid for $\log_{10} (E/\mathrm{eV})=[16.0,21.0]$ and $\log_{10}(-\eta)=[-16,-1]$:
\begin{enumerate}
    \item The standard (LI) values of $\langle N_\mu \rangle$ and $\sigma_{\mu}$ for each primary and hadronic interaction model are fitted as a function of energy using an exponential function (see Eq~\ref{eq:expfunc});
    
    \item The LIV-modified values are expressed through the ratios:
    
    \begin{equation}\label{eq.ratios}
        R_\mu(\epsilon) = \frac{\langle N_\mu \rangle^\text{STD}}{\langle N_\mu \rangle^\eta}, \qquad R_\sigma(\epsilon) = \frac{\sigma^\text{STD}}{\sigma^\eta},
    \end{equation} 
    and parameterised as functions of $\epsilon=\log_{10}(E/\mathrm{eV})$ for each $\eta$. Their fit parameters are subsequently modeled as quadratic functions of $\log_{10}(-\eta)$ (see Eq~\ref{eq:ratios}).
\end{enumerate}

\noindent
This choice of treating separately the LI reference and the LIV-to-standard ratios is motivated by the fact that, once these parameterisations are known for protons and iron, the mean and variance for any mixed mass composition and any value of $\eta$ can be reconstructed a posteriori by combining the proton and iron contributions according to the assumed mass fractions.

\subsection{Parameterisations of Standard Muon Mean and Fluctuations}
\label{sec:A_STD}
The standard (Lorentz–invariant) values of the mean and standard deviation
of the muon number are modelled, as a function of energy, with the same
exponential form:
\begin{equation}
    f(\epsilon) = a \cdot \exp(b \cdot \epsilon)\,,
    \label{eq:expfunc}
\end{equation}
where $\epsilon = \log_{10}(E/\mathrm{eV})$. The coefficients $a$ and $b$ are determined independently for $\langle N_\mu \rangle$ and $\sigma_{N_\mu}$ and for each combination of primary mass and hadronic interaction model, and are reported in Tab.~\ref{tab:eta0_fit}. An  example of this exponential description for EPOS–LHC protons is provided by the $\eta = 0$ curves in the upper–left panel of Fig.~\ref{fig:mean-fit} for the mean
and of Fig.~\ref{fig:rms-fit} for the RMS.

\begin{table}[h]
\centering
\begin{tabular}{ll|cc|cc}
\hline
Model & primary & \multicolumn{2}{c|}{$\langle N_\mu \rangle$} & \multicolumn{2}{c}{$\sigma_{N_\mu}$}\\
& & $a$ [$\times 10^{-11}$] & $b$ & $a$ [$\times 10^{-11}$] & $b$ \\
\hline
EPOS-LHC & proton      & $3.4266 \pm 0.0475$ & $2.1382 \pm 0.0008$ & $4.2845 \pm 0.1227$ & $2.0177 \pm 0.0017$ \\
QGSJetII--04 & proton     & $3.9467 \pm 0.0477$ & $2.1288 \pm 0.0006$ & $7.3410 \pm 0.1808$ & $1.9807 \pm 0.0015$\\
EPOS-LHC & iron        & $5.9547 \pm 0.0199$ & $2.1224 \pm 0.0002$ & $1.7325 \pm 0.0599$ & $1.9995 \pm 0.0018$ \\
QGSJetII--04 & iron        & $6.7664 \pm 0.0257$ & $2.1148 \pm 0.0002$ & $3.8215 \pm 0.1436$ & $1.9587 \pm 0.0020$ \\
\hline
\end{tabular}
\caption{Best-fit parameters $a$ and $b$ for the exponential functions $f(\epsilon) = a \cdot \exp(b \epsilon)$ used to describe $\langle N_\mu \rangle$  and $\sigma_{N_\mu}$  at $\eta = 0$.}
\label{tab:eta0_fit}
\end{table}

\subsection{Parameterisations of LIV-to-Standard Ratios}
\label{sec:B_LIV}

The energy dependence of the LIV-to-standard ratios is described by
simple analytic functions of
$\epsilon = \log_{10}(E/\mathrm{eV})$. For the mean number of muons, the ratio $R_\mu(\epsilon,\eta)$ is parametrised as:
\begin{equation}
    R_\mu(\epsilon,\eta)
    = C(\eta)
      + \frac{1 - C(\eta)}
             {1 + \exp\!\bigl[\lambda(\eta)\,\bigl(\epsilon - \theta(\eta)\bigr)\bigr]}\,,
\end{equation}
where $\theta(\eta)$ sets the transition energy between the
Lorentz-invariant and LIV-dominated regimes, $\lambda(\eta)$ controls the steepness of the transition, and $C(\eta)$ is the high-energy plateau value of the ratio. With this choice $R_\mu \to 1$ at low energies and $R_\mu \to C(\eta)$ at high energies.\\

For the RMS of the muon number, the ratio $R_\sigma(\epsilon,\eta)$ is
modelled as an ``inverted Fermi--Gaussian'' in the energy variable
$\epsilon = \log_{10}(E/\mathrm{eV})$:
\begin{equation}
    R_\sigma(\epsilon,\eta)
    = 1
      + \bigl[1 - f(\epsilon,\eta)\bigr]\,
        \biggl[
            A(\eta)\,
            \exp\!\left(
                -\frac{\bigl(\epsilon - \mu(\eta)\bigr)^2}{2\,s^2}
            \right)
            + \Delta(\eta)
        \biggr],
\end{equation}
with
\begin{equation}
    f(\epsilon,\eta)
    = \frac{1}{1 + \exp\!\bigl[t\,\bigl(\epsilon - \omega(\eta)\bigr)\bigr]}\,,
\end{equation}
where $s = 2$ and $t = 3.2$ are kept fixed. Here $\omega(\eta)$
controls the onset of the transition, $\mu(\eta)$ sets the position of the peak in $\epsilon$, $A(\eta)$ is the Gaussian amplitude and $\Delta(\eta)$ is a constant offset.

\begin{figure}[tb!]
    \centering
    \setlength{\tabcolsep}{6pt}
    \renewcommand{\arraystretch}{1.0}

    \begin{tabular}{c}
        \includegraphics[width=0.9\textwidth]{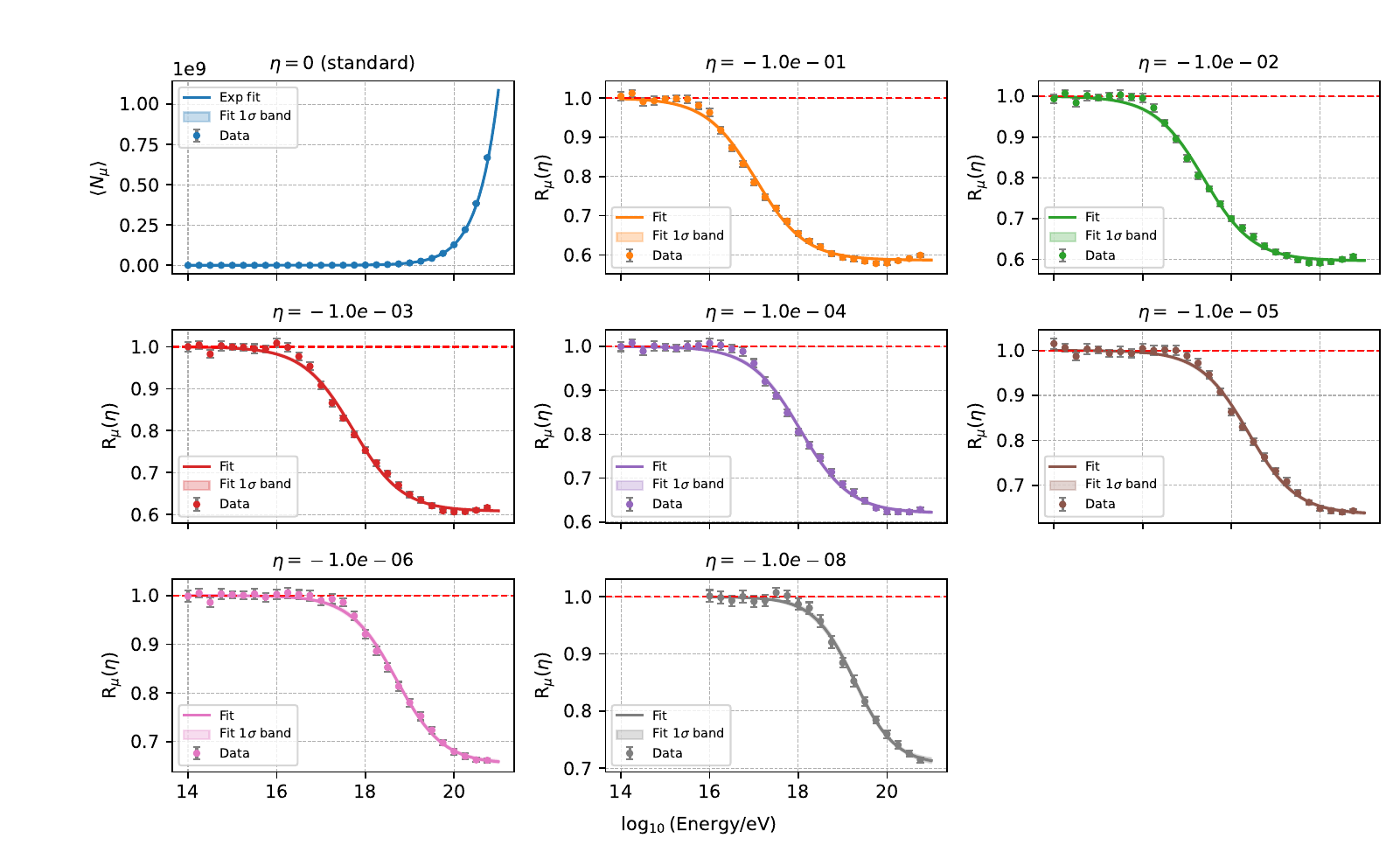} \\[4pt]
        
        \includegraphics[width=0.85\textwidth]{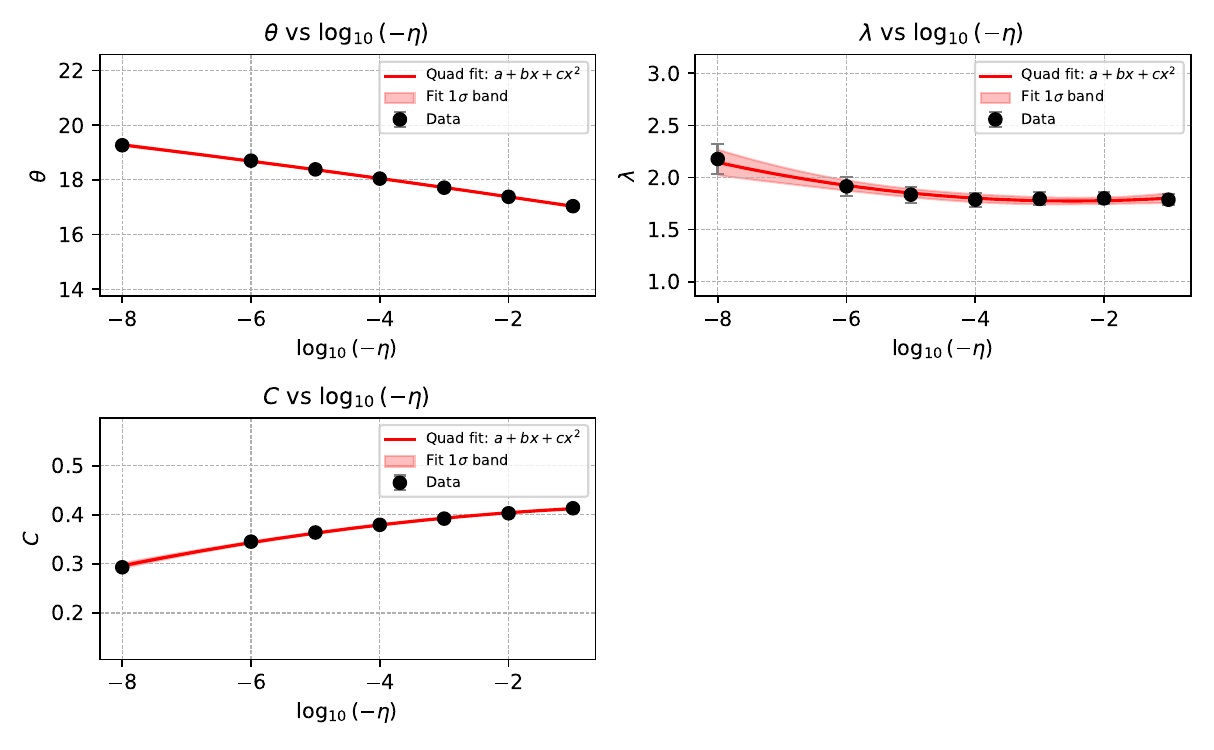} 
     \end{tabular}

    \caption{Mean muon number and LIV-to-standard ratios for EPOS--LHC protons. Top panels: standard case $\eta = 0$ and ratios $R_\mu = \langle N_\mu \rangle^{\mathrm{STD}} / \langle N_\mu \rangle^{\eta}$ as a function of $\log_{10}(E/\mathrm{eV})$ for different $\eta$ values (data points, best–fit curves and $1\sigma$ bands). Bottom panels: corresponding parameters $\theta(\eta)$, $\lambda(\eta)$ and $C(\eta)$ versus $\log_{10}(-\eta)$; points are the results of individual ratio fits, while red curves and bands show the quadratic parameterisations and their $1\sigma$ uncertainties.}
    \label{fig:mean-fit}
\end{figure}

\begin{figure}[tb!]
    \centering
    \setlength{\tabcolsep}{6pt}
    \renewcommand{\arraystretch}{1.0}

    \begin{tabular}{c}
        \includegraphics[width=0.9\textwidth]{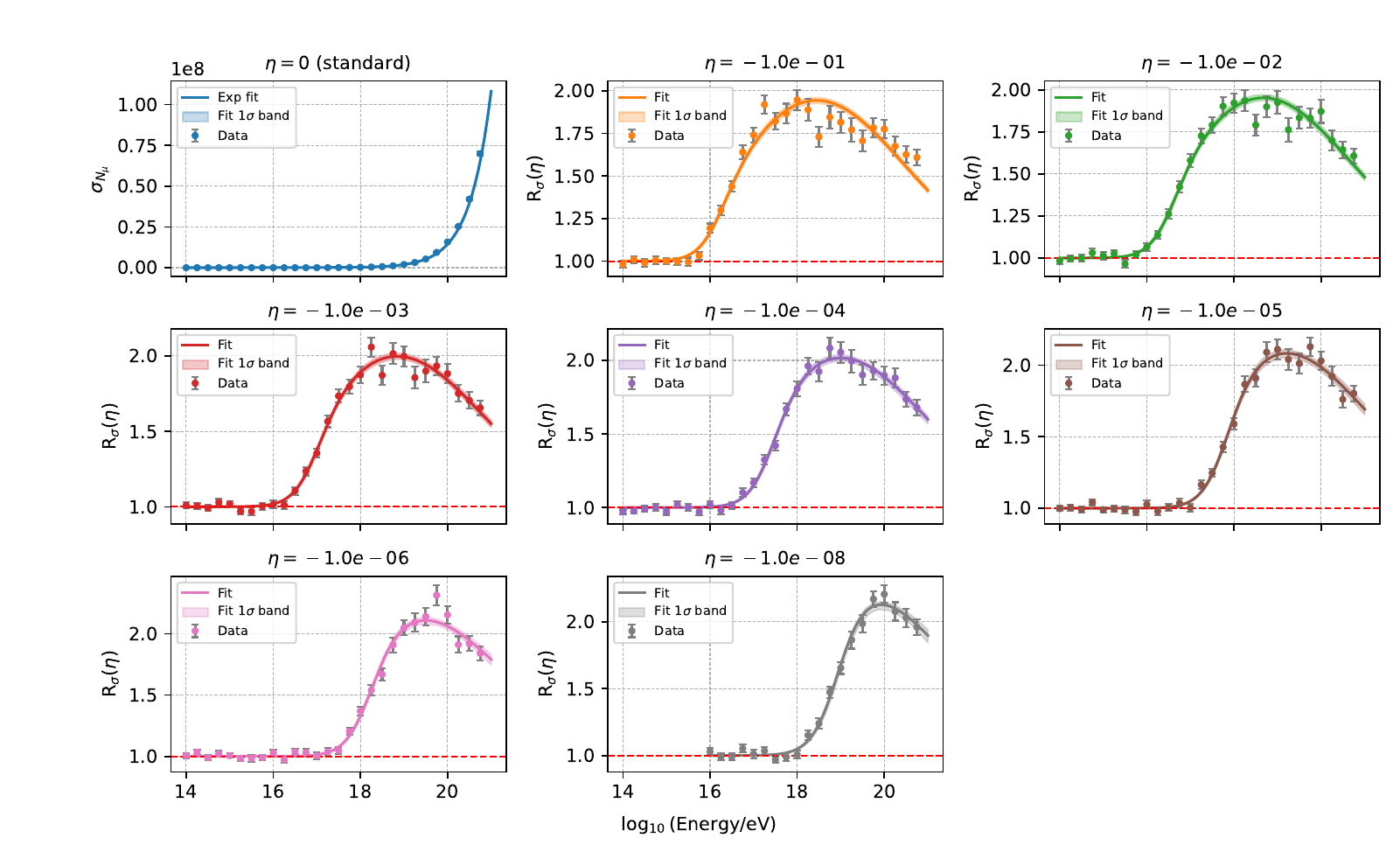}  \\[4pt]
        \includegraphics[width=0.85\textwidth]{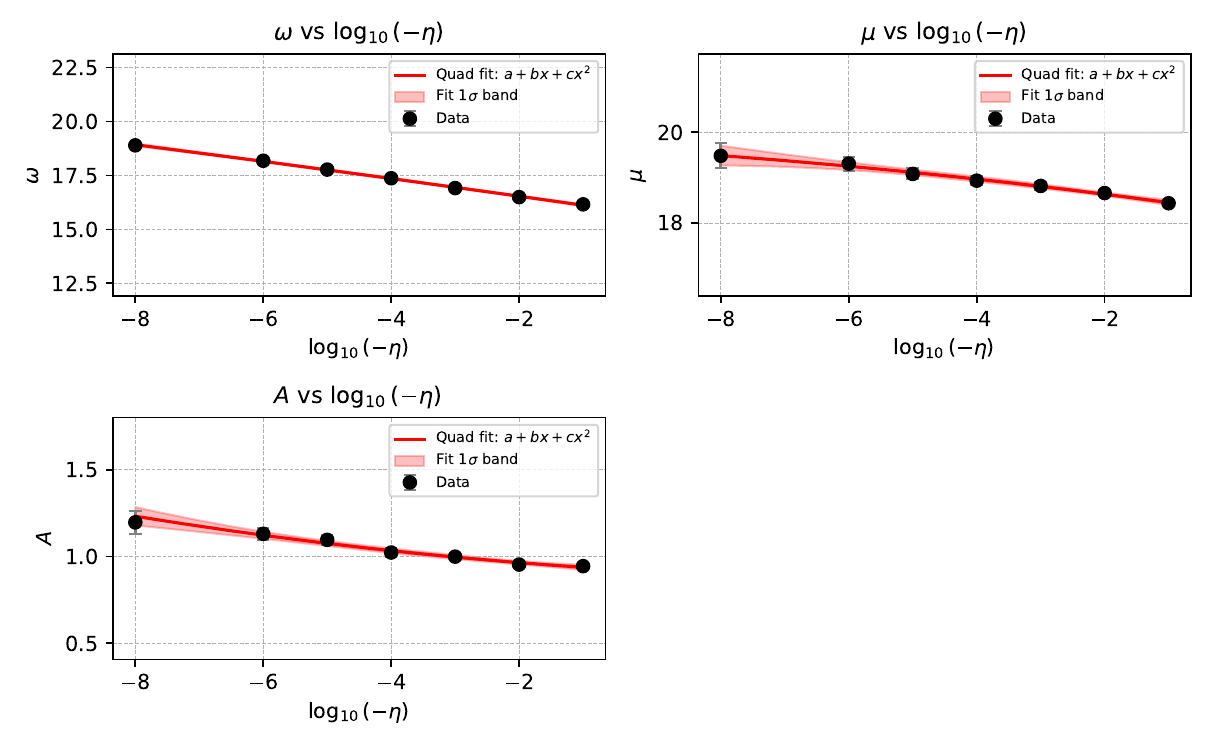}
    \end{tabular}

    \caption{Muon RMS and LIV-to-standard ratios for EPOS--LHC protons. Top panels: standard case $\eta = 0$ and ratios $R_\sigma = \sigma^{\mathrm{STD}} / \sigma^{\eta}$ as a function of $\log_{10}(E/\mathrm{eV})$ for different $\eta$ values (data points, best–fit curves and $1\sigma$ bands). Bottom panels: corresponding parameters $\omega(\eta)$, $\mu(\eta)$ and $N(\eta)$ versus $\log_{10}(-\eta)$; points are the results of individual ratio fits, while red curves and bands show the quadratic parameterisations and their $1\sigma$ uncertainties.}
    \label{fig:rms-fit}
\end{figure}

For protons, we set $\Delta(\eta)=0$ and fit $A(\eta)$, $\mu(\eta)$ and $\omega(\eta)$ as functions of $\log_{10}(-\eta)$. For iron, the data are well described by fixing the Gaussian amplitude to a constant value $A(\eta) \equiv A_0 = 0.4$ and introducing a non-zero high-energy offset $\Delta(\eta) = R_\infty - 1$. In the implementation used in this work we take $R_\infty = 1.0$ for EPOS--LHC and $R_\infty = 0.7$ for QGSJetII--04, so that $R_\sigma(\epsilon,\eta)$ tends to these values at the highest energies. Illustrative examples of these fits for different values of $\eta$, for EPOS--LHC protons and iron, are shown in Fig.~\ref{fig:mean-fit} and Fig.~\ref{fig:rms-fit}, where the upper panels display $R_\mu(\epsilon,\eta)$ and $R_\sigma(\epsilon,\eta)$ as a function of $\epsilon$ for several LIV strengths.

Each parameter $p(\eta) \in \{\theta,\lambda,C,\omega,\mu,A\}$ in the above functions is then parameterised as a function of $\log_{10}(-\eta)$. In general we adopt the quadratic form
\begin{equation}
    p(\eta)
    = a + b\,\log_{10}(-\eta) + c\,\bigl[\log_{10}(-\eta)\bigr]^2\,,
    \label{eq:ratios}
\end{equation}
while for the iron RMS case a linear dependence ($c = 0$) provides an adequate description and is used in the fits. Representative examples of these parameter–vs–$\log_{10}(-\eta)$ fits are shown in the lower panels of Fig.~\ref{fig:mean-fit} and Fig.~\ref{fig:rms-fit}, where the points correspond to individual ratio fits (namely $R_{\mu}$ and $R_{\sigma}$) and the curves to the global parameterisations with their $1\sigma$ uncertainty bands.

\begin{table}[h]
\centering
\resizebox{\textwidth}{!}{%
\begin{tabular}{ll|ccc|ccc|ccc}
\hline
Model & primary & \multicolumn{3}{c|}{$\theta$} & \multicolumn{3}{c|}{$\lambda$} & \multicolumn{3}{c}{$C$} \\
& & $a$ & $b$ & $c$ [$\times 10^{-3}$] & $a$ & $b$ & $c$ & $a$ & $b$ [$\times 10^{-3}$] & $c$ [$\times 10^{-3}$]\\
\hline
EPOS-LHC & proton & $16.67 \pm 0.028$ & $-0.364 \pm 0.017$ & $-4.710 \pm 1.987$ & $1.851 \pm 0.086$ & $0.061 \pm 0.052$ & $0.012 \pm 0.007$ & $0.418 \pm 0.003$ & $4.12 \pm 2.02$ & $-1.39 \pm 0.26$ \\
QGSJetII--04 & proton & $16.82 \pm 0.022$ & $-0.385 \pm 0.013$ & $-7.283 \pm 1.516$ &  $1.740 \pm 0.063$ & $0.032 \pm 0.038$ & $0.011 \pm 0.005$ & $0.430 \pm 0.003$ & $5.63 \pm 1.62$ & $-1.41 \pm 0.22$ \\
EPOS-LHC & iron   & $18.38 \pm 0.005$ & $-0.353 \pm 0.004$ & $-11.330 \pm 0.514$ &  $1.659 \pm 0.015$ & $0.063 \pm 0.011$ & $0.030 \pm 0.002$ & $0.445 \pm 0.001$ & $6.24 \pm 0.80$ & $-2.98 \pm 0.18$ \\
QGSJetII--04 & iron   & $18.48 \pm 0.005$ & $-0.363 \pm 0.004$ & $-11.202 \pm 0.641$ &  $1.634 \pm 0.017$ & $0.003 \pm 0.010$ & $0.022 \pm 0.002$ & $0.445 \pm 0.001$ & $9.23 \pm 0.96$ & $-2.75 \pm 0.15$ \\
\hline
\end{tabular}
}
\caption{Fit parameters for $R_\mu$ as a function of $\log_{10}(-\eta)$: $\theta$, $\lambda$, and $C$.}
\label{tab:ratio_mean}
\end{table}

\vspace{1em}
\begin{table}[ht]
\centering
\resizebox{\textwidth}{!}{
\begin{tabular}{ll|ccc|ccc|ccc}
\hline
Model & primary & \multicolumn{3}{c|}{$\omega$} & \multicolumn{3}{c|}{$\mu$} & \multicolumn{3}{c}{$A$} \\
& & $a$ & $b$ & $c$ [$\times 10^{-3}$] & $a$ & $b$ & $c$ [$\times 10^{-3}$] & $a$ & $b$ & $c$ [$\times 10^{-3}$]\\
\hline
EPOS-LHC & proton & $15.694 \pm 0.076$ & $-0.428 \pm 0.040$ & $-2.982 \pm 4.505$ & $18.252 \pm 0.116$ & $-0.204 \pm 0.077$ & $-6.225 \pm 10.628$ & $0.916 \pm 0.030$ & $-0.019 \pm 0.018$ & $2.535 \pm 2.633$\\
QGSJetII--04 & proton & $15.605 \pm 0.088$ & $-0.580 \pm 0.046$ & $-2.508\pm 5.297$ & $18.381 \pm 0.140$ & $0.124 \pm 0.088$ & $51.680 \pm 0.002$ & $0.576 \pm 0.025$ & $-0.044\pm 0.016$ & $-1.025\pm2.117$ \\
EPOS-LHC & iron   & $17.504 \pm 0.048$ & $-0.330 \pm 0.014$& -- & $15.861 \pm 0.091$ & $-0.591 \pm 0.027$ & --& $0.4$ & -- & --\\

QGSJetII--04 & iron   & $17.632 \pm 0.129$ & $-0.289 \pm 0.028$ & -- & $16.949 \pm 0.057$ & $-0.424 \pm 0.017$ & -- & $0.4$ & -- & -- \\
\hline
\end{tabular}
}
\caption{Fit parameters for $R_\sigma$ as a function of $\log_{10}(-\eta)$: $\omega$, $\mu$ and $A$.}
\label{tab:ratio_rms}
\end{table}

\section{III.~Composition Maximizing Relative Fluctuations in LIV Scenarios}
\begin{figure}[tb!]
    \centering
    \setlength{\tabcolsep}{6pt}
    \renewcommand{\arraystretch}{1.0}

    \begin{tabular}{cc}
        \includegraphics[width=0.45\textwidth]{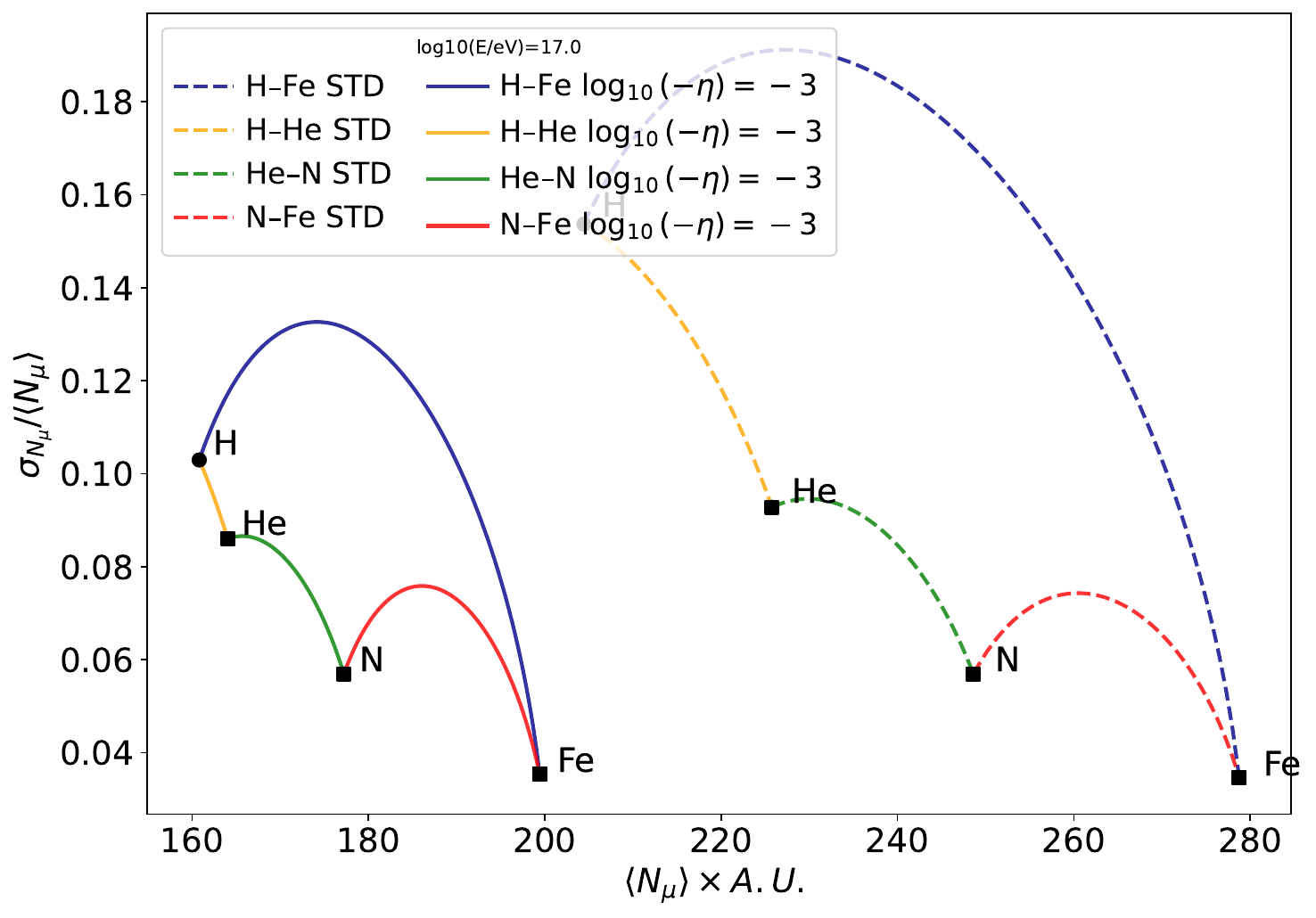} &
        \includegraphics[width=0.45\textwidth]{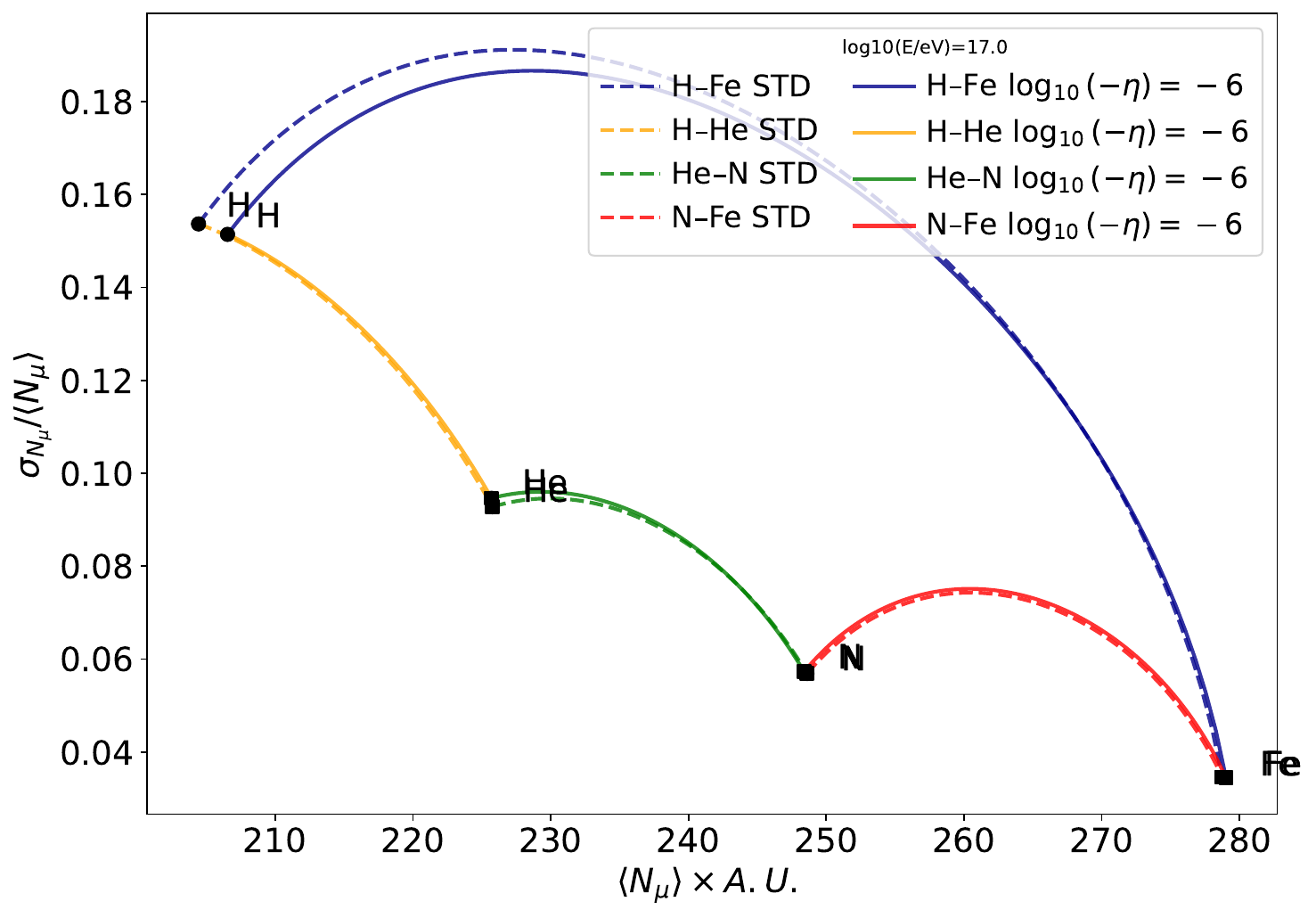} \\[4pt]

        \includegraphics[width=0.45\textwidth]{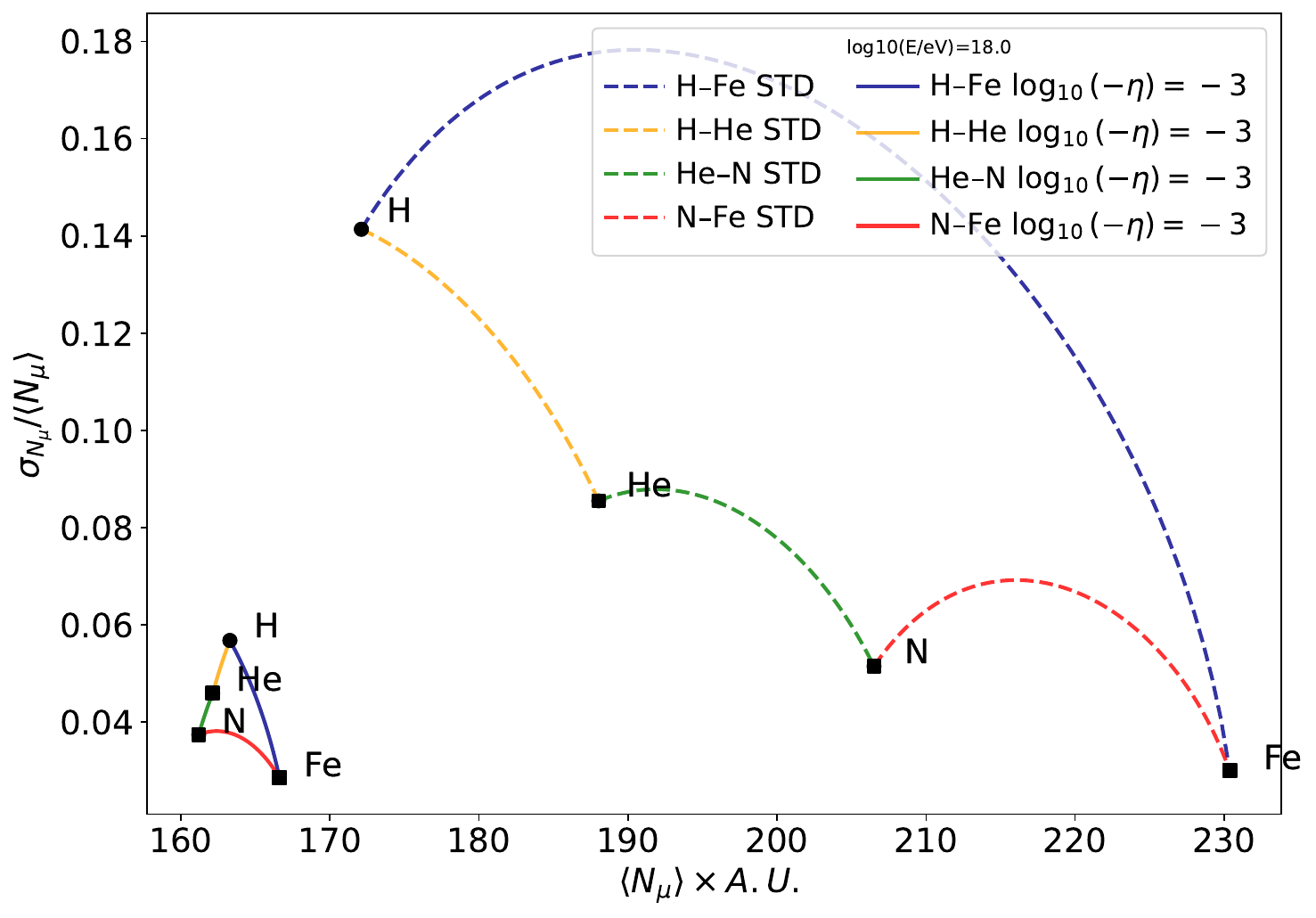} &
        \includegraphics[width=0.45\textwidth]{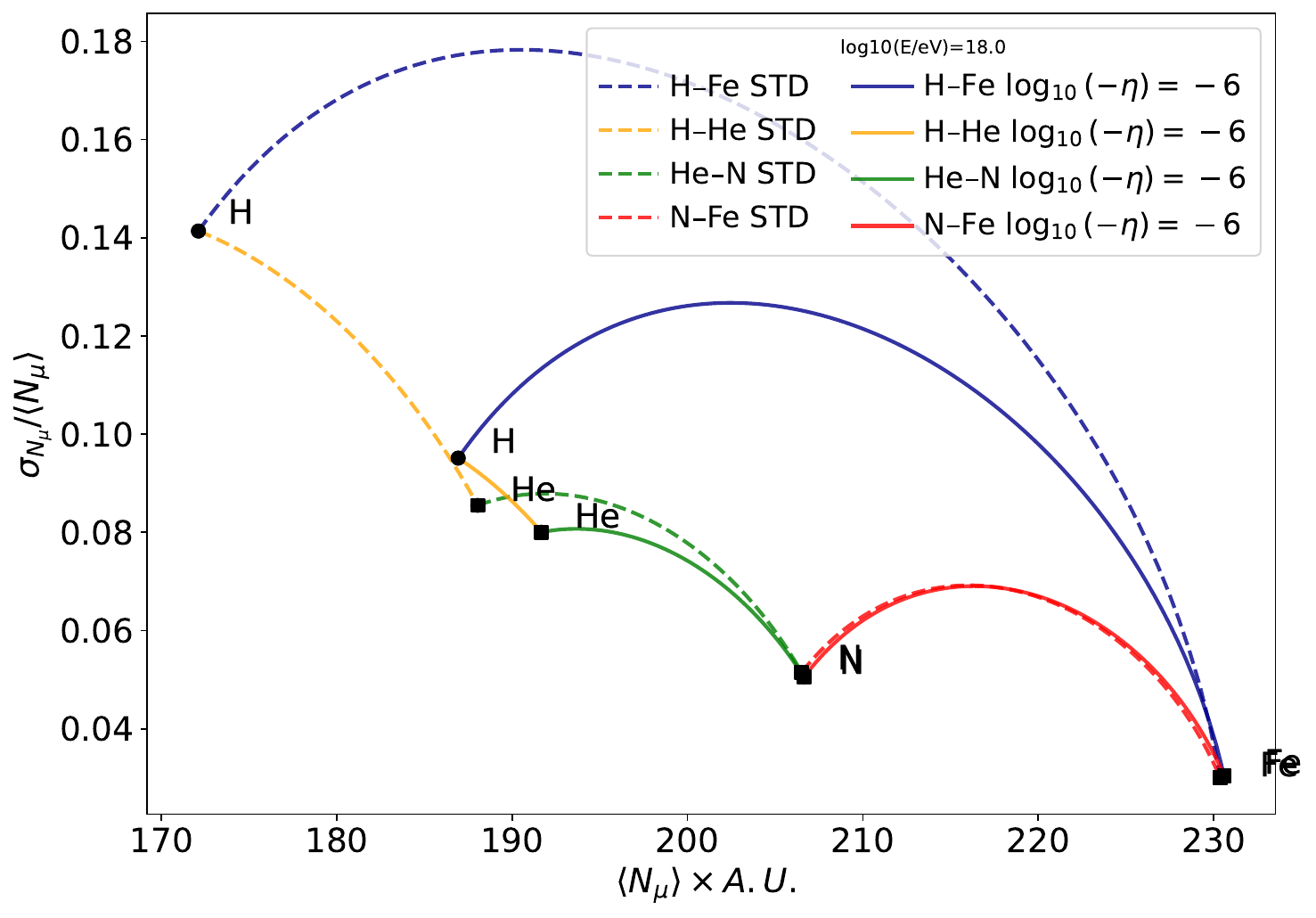} \\[4pt]

        \includegraphics[width=0.45\textwidth]{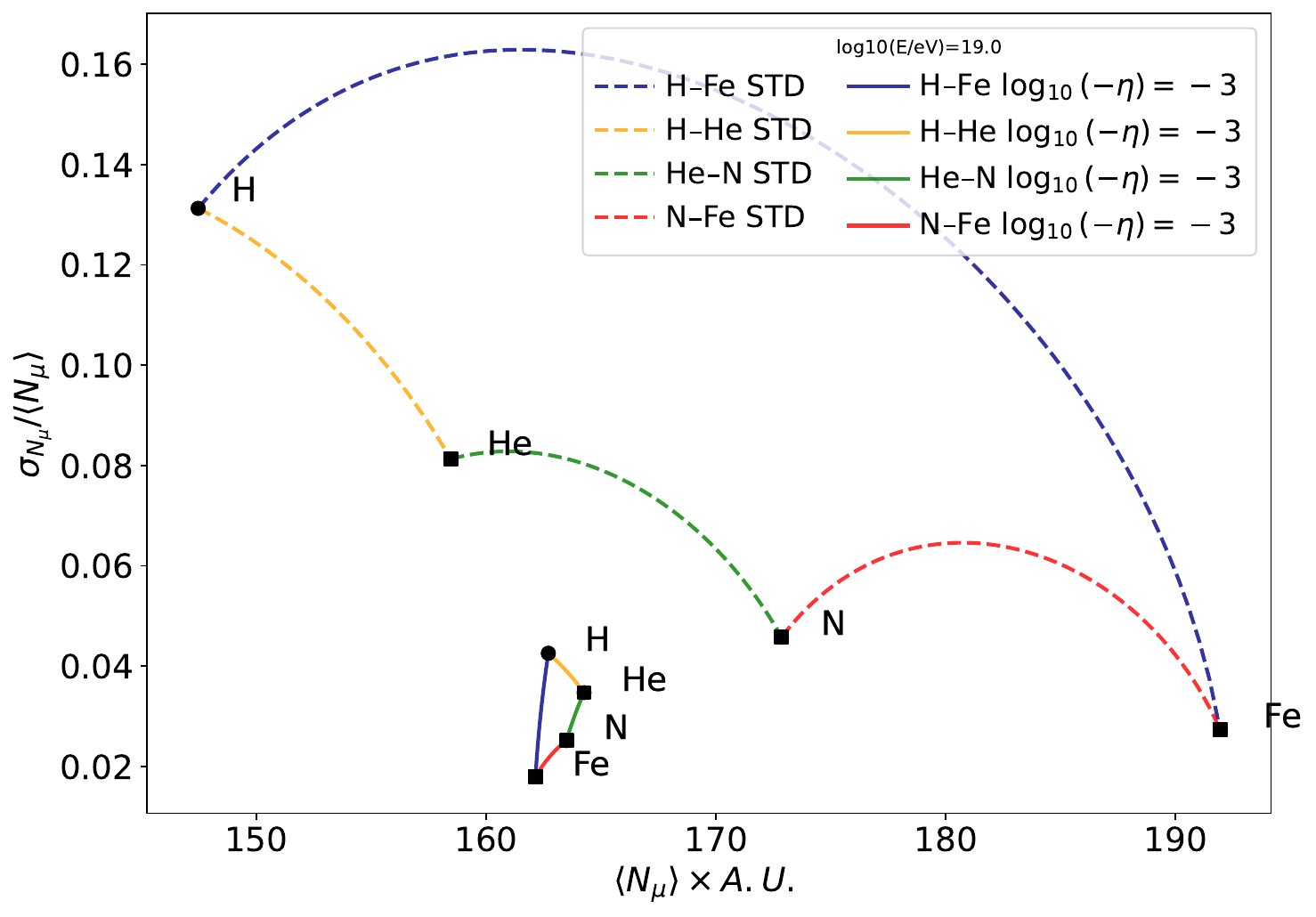} &
        \includegraphics[width=0.45\textwidth]{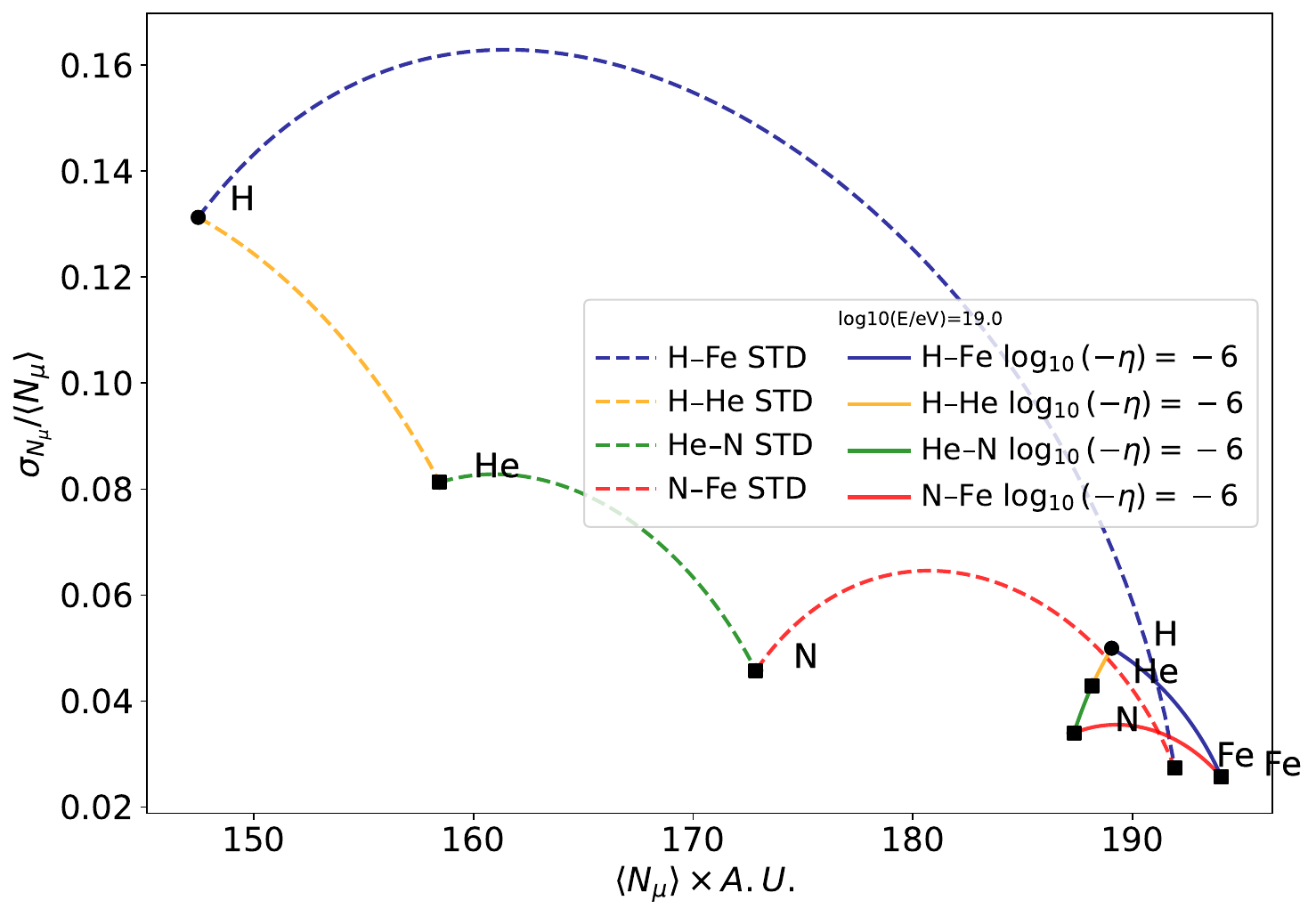}
    \end{tabular}

    \caption{Relative muon fluctuations $\sigma_{N_\mu} / \langle N_\mu \rangle$ as a function of $\langle N_\mu \rangle$ for selected primary combinations at three energies $\log_{10}(E/\text{eV}) = 17, 18, 19$, for the EPOS-LHC hadronic interaction model. Left column: $\log_{10}(-\eta)=3$, corresponding to a large LIV violation. Right column: $\log_{10}(-\eta)=6$, corresponding to the region currently constrained by this analysis.}
    \label{fig:umbrella-panel}
\end{figure}

In order to identify the most conservative scenario for constraining LIV effects through muon fluctuations, we investigated how the relative muon fluctuations $\sigma_{N_\mu}/\langle N_\mu \rangle$ depend on mass composition for different values of energy and the LIV parameter $\eta$.

Figures~\ref{fig:umbrella-panel} shows “umbrella plots” of relative muon fluctuations at fixed values of $\log_{10}(-\eta) = -3$ and $-6$, corresponding respectively to a large LIV violation and to the region currently constrained by this analysis. Each panel shows $\sigma_{N_\mu} / \langle N_\mu \rangle$ as a function of $\langle N_\mu \rangle$ for selected primary combinations at three energies in the relevant range, with the corresponding standard (non‑LIV) case superimposed for comparison.
Despite the distortions in shower development induced by LIV, the mass hierarchy in muon production is largely preserved. This implies that the composition that maximizes the relative fluctuation can generally achieved with a binary mixture of protons and iron. In some extreme configurations (large $\eta$ and/or high energy), the maximum is obtained for pure proton composition, but never for intermediate-mass combinations. 

\begin{figure*}[t]
    \centering
    \begin{minipage}{0.49\textwidth}
        \centering
        \includegraphics[width=\linewidth]{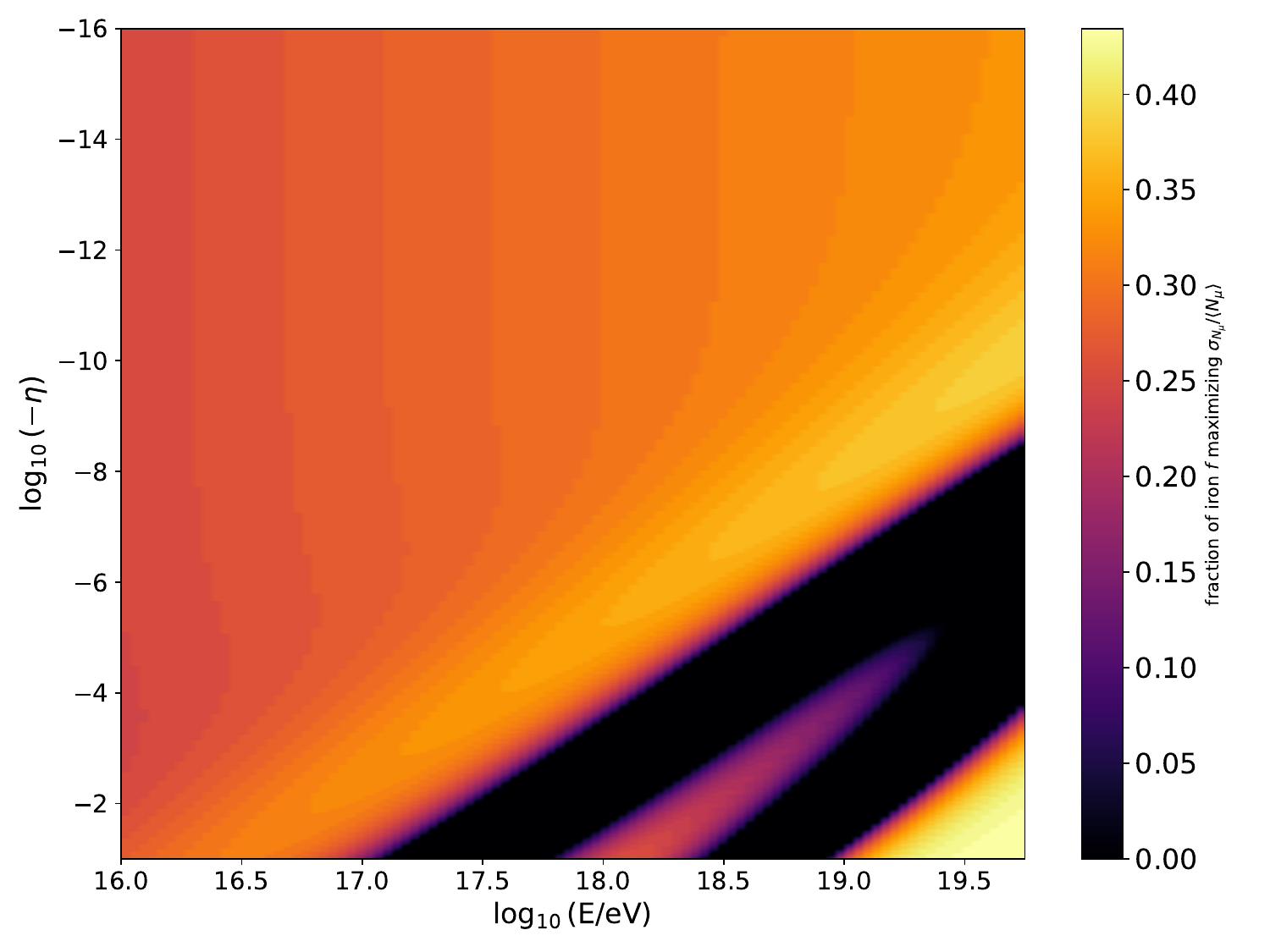}
    \end{minipage}
    \hfill
    \begin{minipage}{0.49\textwidth}
        \centering
        \includegraphics[width=\linewidth]{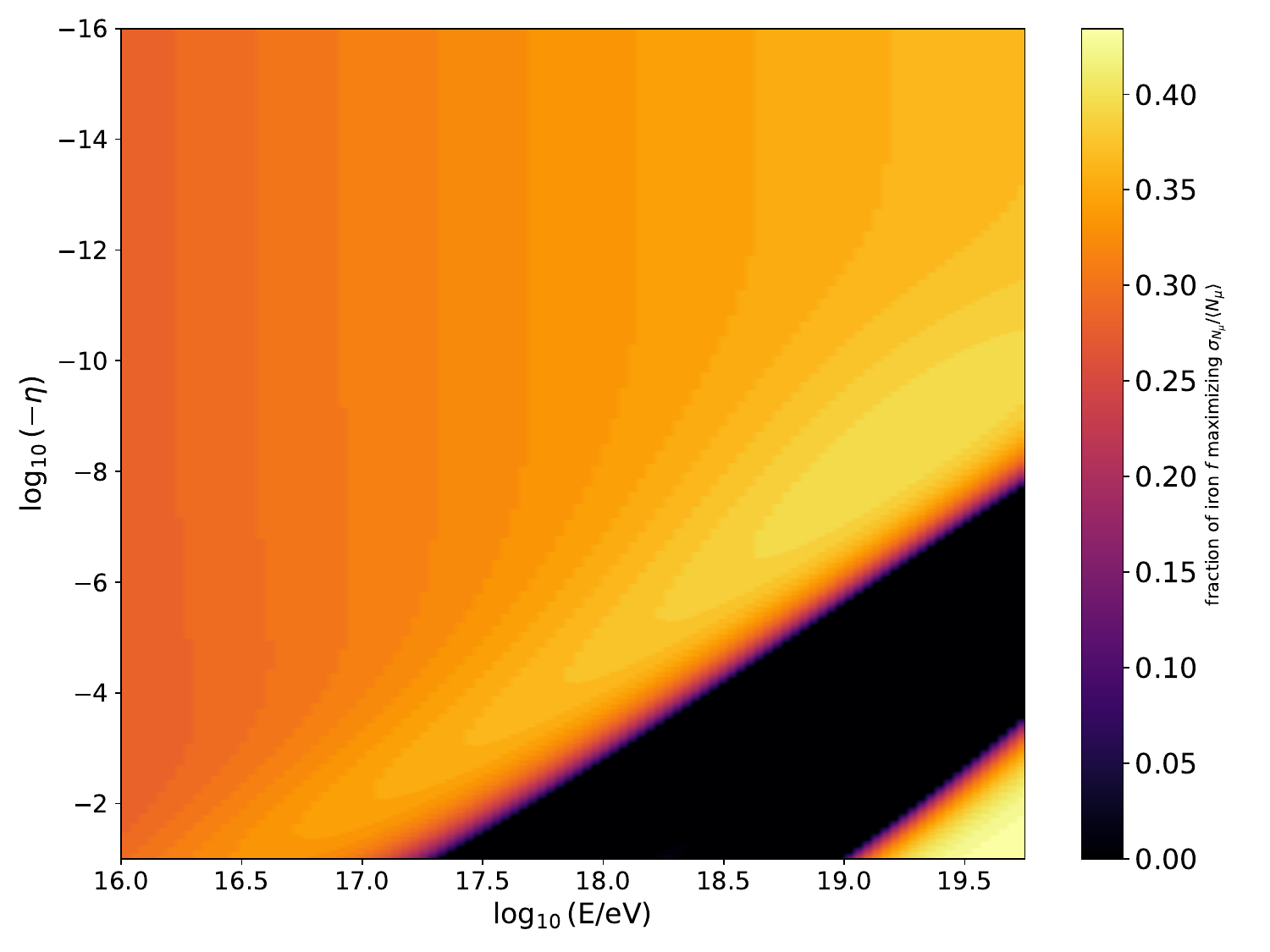}
    \end{minipage}

    \caption{Fraction of iron $f$ that maximizes the relative muon fluctuations $\sigma_{N_\mu} / \langle N_\mu \rangle$, as a function of $\log_{10}(E/\text{eV})$ and $\log_{10}(-\eta)$. The corresponding proton fraction is $1 - f$ for the EPOS-LHC (left panel) and QGSJetII--04 (right panel) hadronic interaction models. }
    \label{fig:max-panel}
\end{figure*}

To identify the actual mass mixture that maximizes the relative muon fluctuations over energy and LIV parameter values, we performed a full scan over binary mixtures of protons and iron, varying the iron fraction $f$ from 1 (pure Fe) to 0 (pure p).
The complete results of this optimization are shown in Fig.~\ref{fig:max-panel}, where the optimal iron fraction $f$ is mapped over the full $(E,\eta)$ space, for the EPOS-LHC (left panel) and QGSJetII-04 (right panel) interaction models. These heatmaps allow the identification of the most conservative scenario for any given configuration, based on the parametrizations derived above, without requiring additional simulations.

\end{document}